\documentclass{aa}
\usepackage{subfigure}
\usepackage[varg]{txfonts}
\usepackage{natbib,twoopt}
\usepackage[breaklinks=true]{hyperref}
\usepackage{graphicx,multicol}
\usepackage{amsmath}
\usepackage{amssymb}
\usepackage{rotating}
\bibpunct{(}{)}{;}{a}{}{,}
\makeatletter
  \newcommandtwoopt{\citeads}[3][][]{\href{http://adsabs.harvard.edu/abs/#3}%
    {\def\hyper@linkstart##1##2{}%
     \let\hyper@linkend\@empty\citealp[#1][#2]{#3}}}
  \newcommandtwoopt{\citepads}[3][][]{\href{http://adsabs.harvard.edu/abs/#3}%
    {\def\hyper@linkstart##1##2{}%
     \let\hyper@linkend\@empty\citep[#1][#2]{#3}}}
  \newcommandtwoopt{\citetads}[3][][]{\href{http://adsabs.harvard.edu/abs/#3}%
    {\def\hyper@linkstart##1##2{}%
     \let\hyper@linkend\@empty\citet[#1][#2]{#3}}}
  \newcommandtwoopt{\citeyearads}[3][][]%
    {\href{http://adsabs.harvard.edu/abs/#3}
    {\def\hyper@linkstart##1##2{}%
     \let\hyper@linkend\@empty\citeyear[#1][#2]{#3}}}
\makeatother

\def\i20210n{{Capturing dual AGN activity and kiloparsec-scale outflows\\in IRAS 20210+1121}}
\def\oii{{[O\,{\scriptsize II}]}}
\def\hb{{H$\beta$}}
\def\oiii{{[O\,{\scriptsize III}]}}
\def\oi{{[O\,{\scriptsize I}]}}
\def\ha{{H$\alpha$}}
\def\n2{{[N\,{\scriptsize II}]}}
\def\s2{{[S\,{\scriptsize II}]}}
\def\h0{{$H_0 = 70$ km s$^{-1}$ Mpc$^{-1}$}}
\def\om{{$\Omega_{\rm M} = 0.3$}}
\def\ol{{$\Omega_\Lambda = 0.7$}}

\title{\i20210n}
\subtitle{}

\author{{F. G. Saturni}\inst{\ref{inst1},\ref{inst2}}
\and
{G. Vietri}\inst{\ref{inst1},\ref{inst3}}
\and
{E. Piconcelli}\inst{\ref{inst1}}
\and
{C. Vignali}\inst{\ref{inst4},\ref{inst5}}
\and
{M. Bischetti}\inst{\ref{inst6}}
\and
{A. Bongiorno}\inst{\ref{inst1}}
\and
{S. Cazzoli}\inst{\ref{inst7}}
\and
{C. Feruglio}\inst{\ref{inst6}}
\and\\
{F. Fiore}\inst{\ref{inst6}}
\and
{B. Husemann}\inst{\ref{inst8}}
\and
{C. Ramos Almeida}\inst{\ref{inst9},\ref{inst10}}
}
\institute{INAF -- Osservatorio Astronomico di Roma, Via Frascati 33, I-00078 Monte Porzio Catone (RM), Italy.\\
\email{francesco.saturni@inaf.it}\label{inst1}
\and
ASI -- Space Science Data Center, Via del Politecnico snc, I-00133 Roma, Italy.\label{inst2}
\and
INAF -- Istituto di Astrofisica Spaziale e Fisica Cosmica di Milano, Via A. Corti 12, I-20133 Milano, Italy.\label{inst3}
\and
Universit{\`a} di Bologna, Dip. di Fisica e Astronomia ``A. Righi'', Via P. Gobetti 93/2, I-40129 Bologna, Italy.\label{inst4}
\and
INAF -- Osservatorio di Astrofisica e Scienza dello Spazio di Bologna, Via P. Gobetti 93/3, I-40129 Bologna, Italy.\label{inst5}
\and
INAF -- Osservatorio Astronomico di Trieste, Via G. B. Tiepolo 11, I-34143 Trieste, Italy.\label{inst6}
\and
CSIC -- Instituto de Astrof{\'i}sica de Andaluc{\'i}a, Dep.to de Astronom{\'i}a Extragal{\'a}ctica, Glorieta de la Astronom{\'i}a s/n, E-18008 Granada, Spain.\label{inst7}
\and
Max-Planck Institut f{\"u}r Astronomie, K{\"o}nigstuhl 17, D-69117 Heidelberg, Germany.\label{inst8}
\and
Instituto de Astrof{\'i}sica de Canarias, C/ V{\'i}a L{\'a}ctea s/n, E-38205 La Laguna (Tenerife), Spain.\label{inst9}
\and
Universidad de La Laguna, Dep.to de Astrof{\'i}sica, Av.da Astrof{\'i}sico F. S{\'a}nchez s/n, E-38206 La Laguna (Tenerife), Spain.\label{inst10}
}
\date{Received 2021 May 24 / Accepted 2021 Jul 26}

\abstract
    {The most standard scenario for the evolution of massive galaxies across cosmic time assumes a correspondence based on the interplay between active galactic nuclei (AGN) feedback, which injects large amounts of energy into the host environment, and galaxy mergers, with their ability to trigger massive star formation events and accretion onto supermassive black holes. Interacting systems hosting AGN are useful laboratories for obtaining key insights into both phenomena. In this context, we present an analysis of the optical spectral properties of IRAS 20210+1121 (I20210), a merging system at $z = 0.056$. According to X-ray data, this object comprises two interacting galaxies, each hosting an obscured AGN. The optical spectra confirm the presence of AGN features in both galaxies. In particular, we are able to provide a Seyfert classification for I20210 North. The spectrum of I20120 South shows broad blueshifted components associated with the most intense emission lines that indicate the presence of an ionized outflow, for which we derive a maximum velocity of $\sim$2000 km s$^{-1}$, an extension of $\sim$2 kpc, and a mass rate of $\sim$0.6 M$_\odot$ yr$^{-1}$. We also report the existence of an ionized nebular component with $v \sim 1000$ km s$^{-1}$ at $\sim$6.5 kpc southwards of I20210 South, which can be interpreted as disrupted gas ejected from the host galaxy by the action of the outflow. I20120 therefore exhibits a double obscured AGN, with one of them showing evidence of ongoing events for AGN-powered outflows. Future spatially resolved spectroscopy will allow for an accurate mapping of the gas kinematics in this AGN pair and evaluate the impact of the outflow on both the interstellar medium and the galaxy environment.}

\keywords{
galaxies: active -- galaxies: groups: general -- galaxies: groups: individual: IRAS 20210+1121 -- galaxies: Seyfert -- quasars: emission lines -- quasars: supermassive black holes
}

\begin{document}

\titlerunning{Dual AGN activity and kpc-scale outflows in IRAS 20210+1121}
\authorrunning{F. G. Saturni et al.}
\maketitle

\section{Introduction }\label{sec:intro}

The past history of formation and evolution of present-day massive galaxies is a key point  to consider on the path to obtaining a fuller understanding of the functioning of the Universe. In this context, the study of processes operating on galaxy-wide scales, such as the presence of active galactic nuclei \citep[AGN; e.g.,][]{Lyn69} or events related to galaxy mergers \citep[e.g.,][]{Her89}, is crucial to improving our knowledge of the mechanisms that are able to boot, maintain, enhance, and quench star formation in galaxies -- thereby shaping the entire environment in the process.

It is now widely accepted that AGN activity and galaxy mergers are among the most effective phenomena regulating star formation in massive galaxies at nearly all redshifts; namely, the first injects large amounts of energy that are able to originate powerful gas winds in the surrounding environment \citep[e.g.,][]{DiM05,Cat09,Fab12}, while the second takes place by triggering massive star formation and starburst events in molecular gas-rich clouds \citep[e.g.,][]{San96}. Both simulations of the evolutionary history of the Universe in the framework of the $\Lambda$-CDM model \citep[e.g.,][]{Dav85,Spr05,Cro06} and observations that confirm their contribution to simultaneously shaping galactic environments  \citep[e.g.,][]{San88,Kor13,Ell19} have confirmed the major role that such processes play in galaxy formation and evolution.

\begin{figure*}[htbp]
\centering
\includegraphics[width=.95\linewidth,angle=90]{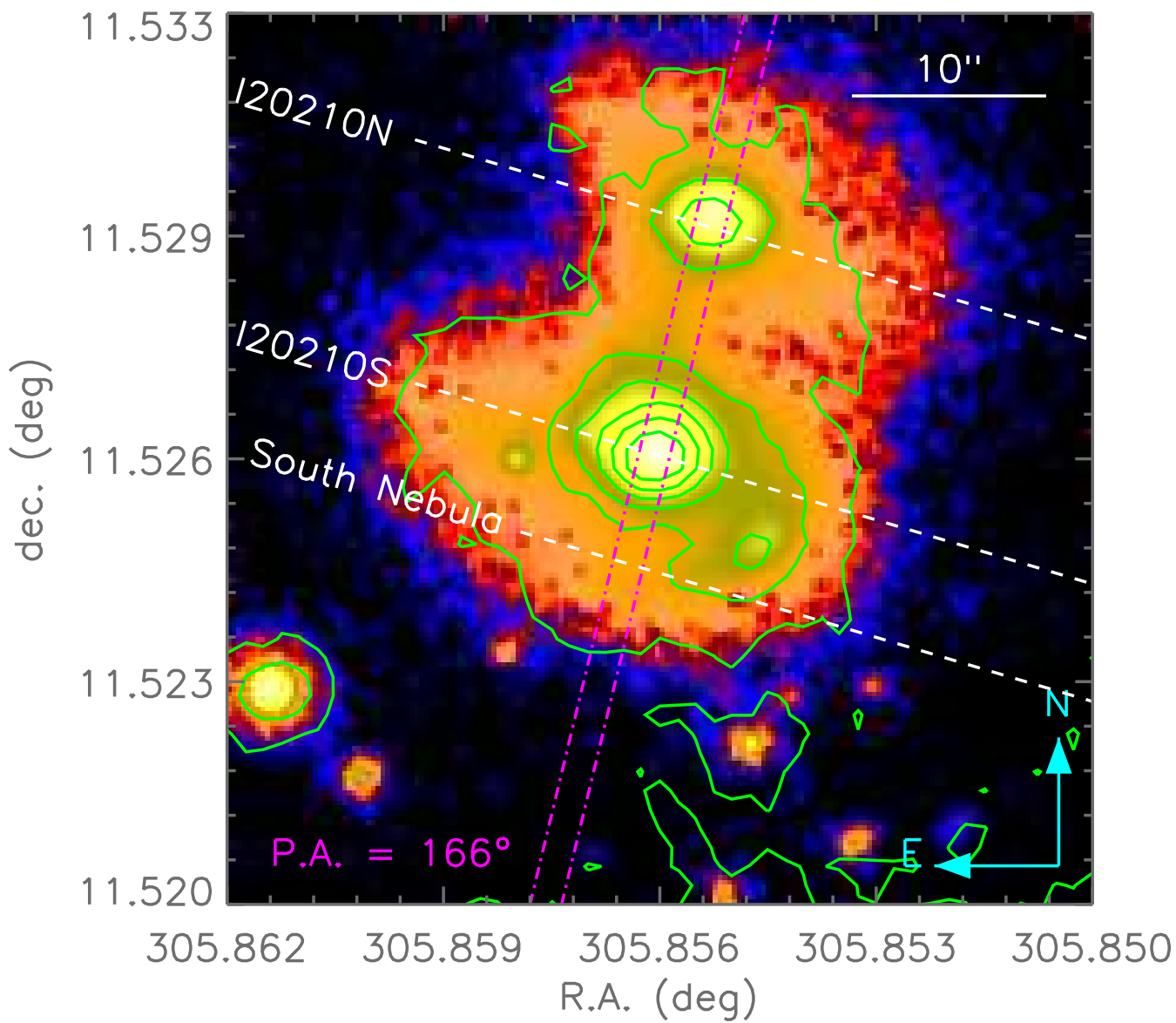}
\caption{Image in false colors of the I20210 system, obtained by combining the {\itshape grizy} exposures of the Pan-STARRS1 survey \citep[PS1;][]{Cha16} centered on the sky coordinates of I20210S ($\alpha_{\rm J2000} = $ 20 23 25.4, $\delta_{\rm J2000} = +$11 31 34.7). The isophotes of the XMM-{\itshape Newton} Optical Monitor (OM) UVW1 mosaic exposure (green solid lines) taken simultaneously to the X-ray data analyzed by \citet{Pic10} -- along with some of the associated CCD count levels -- are drawn onto the PS1 image to highlight weak features. The TNG slit direction and position (magenta dot-dashed lines) are also indicated along with the positions and directions of the trace centers (white dashed lines) identified to extract the 1D spectrum of each object.}
\label{fig:photimg}
\end{figure*}

Within this general picture, however, several values related to how exactly AGN energetics and mergers directly impact the star formation history of galaxies are still missing. For instance, it is still unknown whether radiation-powered gas outflows are ubiquitous to all AGN \citep[e.g.,][]{Elv00} or whether they affect only a fraction of the AGN lifetime \citep[e.g.,][]{Far07}, along with what their effectiveness is with regard to altering the physical and dynamical status of gas reservoirs on several spatial scales \citep[e.g.,][]{Sca04,Cic18}. In addition, the relative dominance of one process onto the other for moving large gas masses and triggering or quenching star formation has been found to be dependent on the details of the AGN emission mode, the galaxy's surrounding environment and its star formation history \citep[e.g.,][]{Hop06,Hec14}. Therefore, the study of interacting galactic systems with the presence of multiple AGN \citep[e.g.,][]{Vei02} offers an extremely interesting possibility for understanding the properties and links between such competing mechanisms. The most common objects of this kind are dual AGN, in which an active nucleus is hosted in both members of a pair of interacting galaxies with separation on the scale of $5 \div 20$ kpc \citep[see e.g.,][and references therein]{DeR18}.

\begin{figure*}[htbp]
\centering
\begin{minipage}{.49\textwidth}
\includegraphics[scale=0.45]{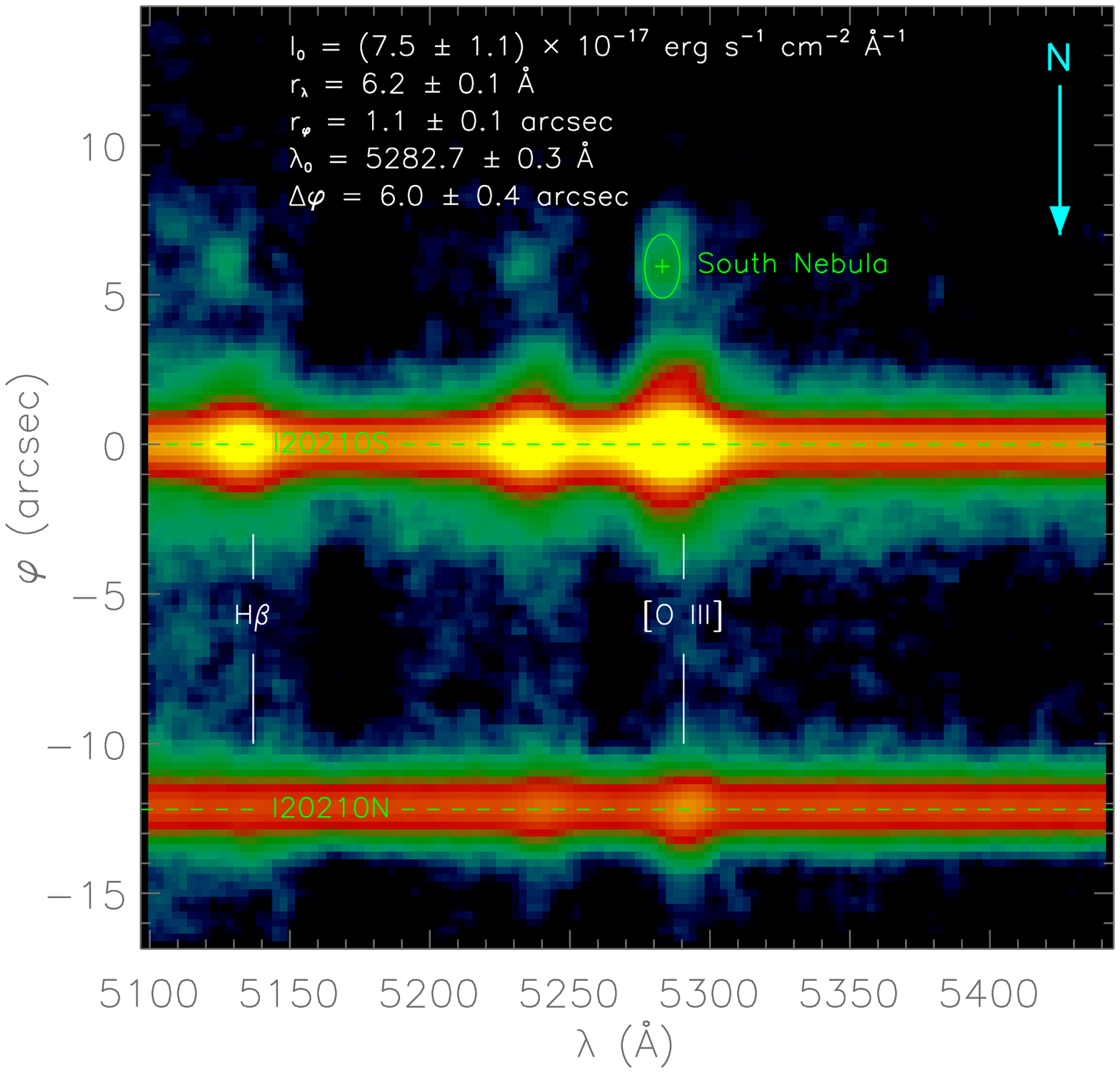}
\end{minipage}
\begin{minipage}{.49\textwidth}
\includegraphics[scale=0.45]{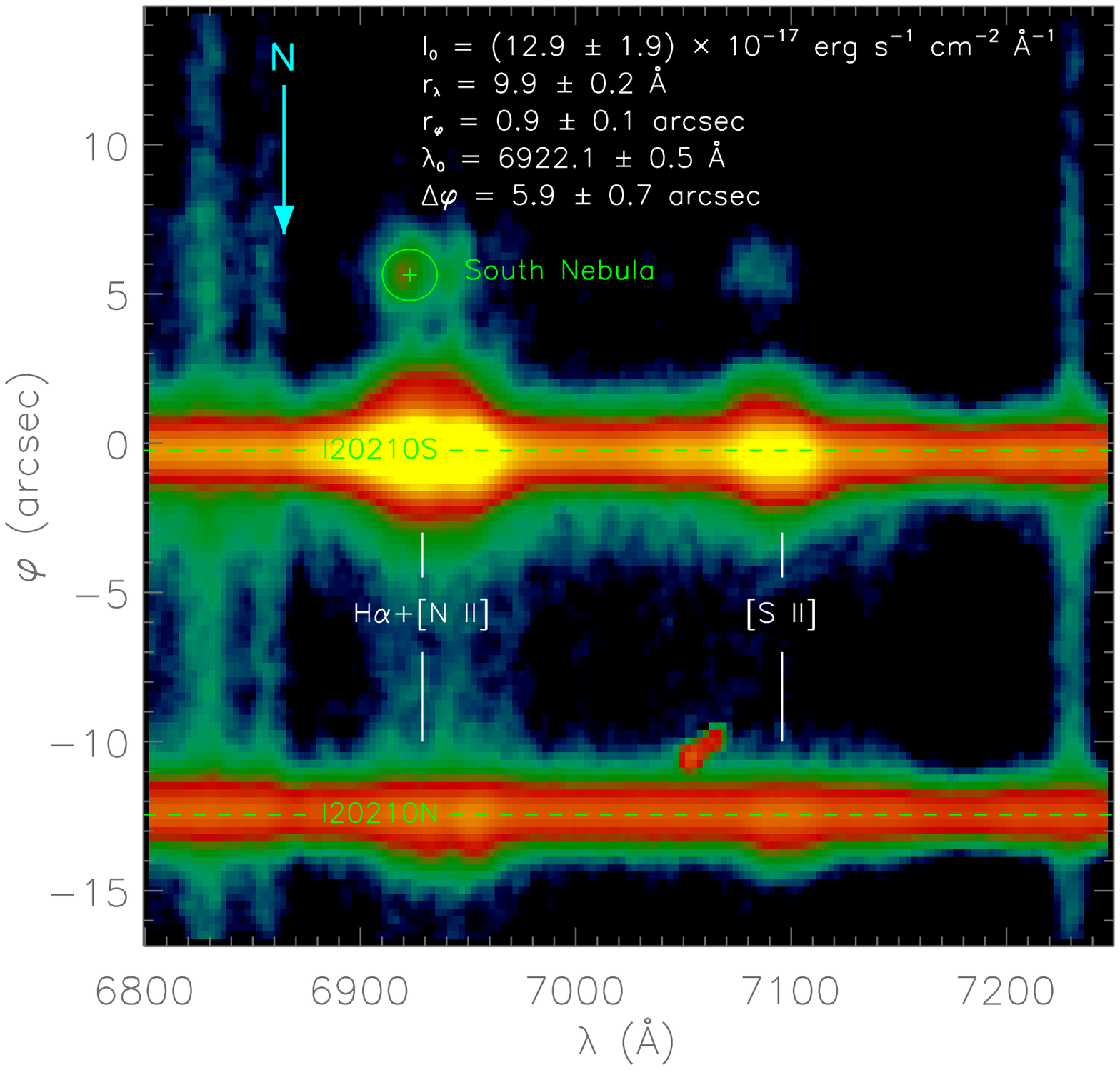}
\end{minipage}
\caption{2D spectral sections of the I20210 system. {\itshape Left panel:} The \hb\ and \oiii\ region. {\itshape Right panel:} The \ha+\n2\ and \s2\ region. In both panels, the trace centers of the main components are identified for reference (green dashed lines). The elliptical fits to the \oiii\ and \ha+\n2\ emissions from the South Nebula are reported (green ellipses) along with the respective best-fit parameters and statistical uncertainties. In the right panel, the vertical features are sky lines, whereas the extrusion close to the \s2\ emission of I20210N is a cluster of saturated pixels that is excluded from the {\scriptsize IRAF} extraction of the 1D spectrum.}
\label{fig:blobs}
\end{figure*}

\begin{figure*}[htbp]
\centering
\includegraphics[width=.95\linewidth]{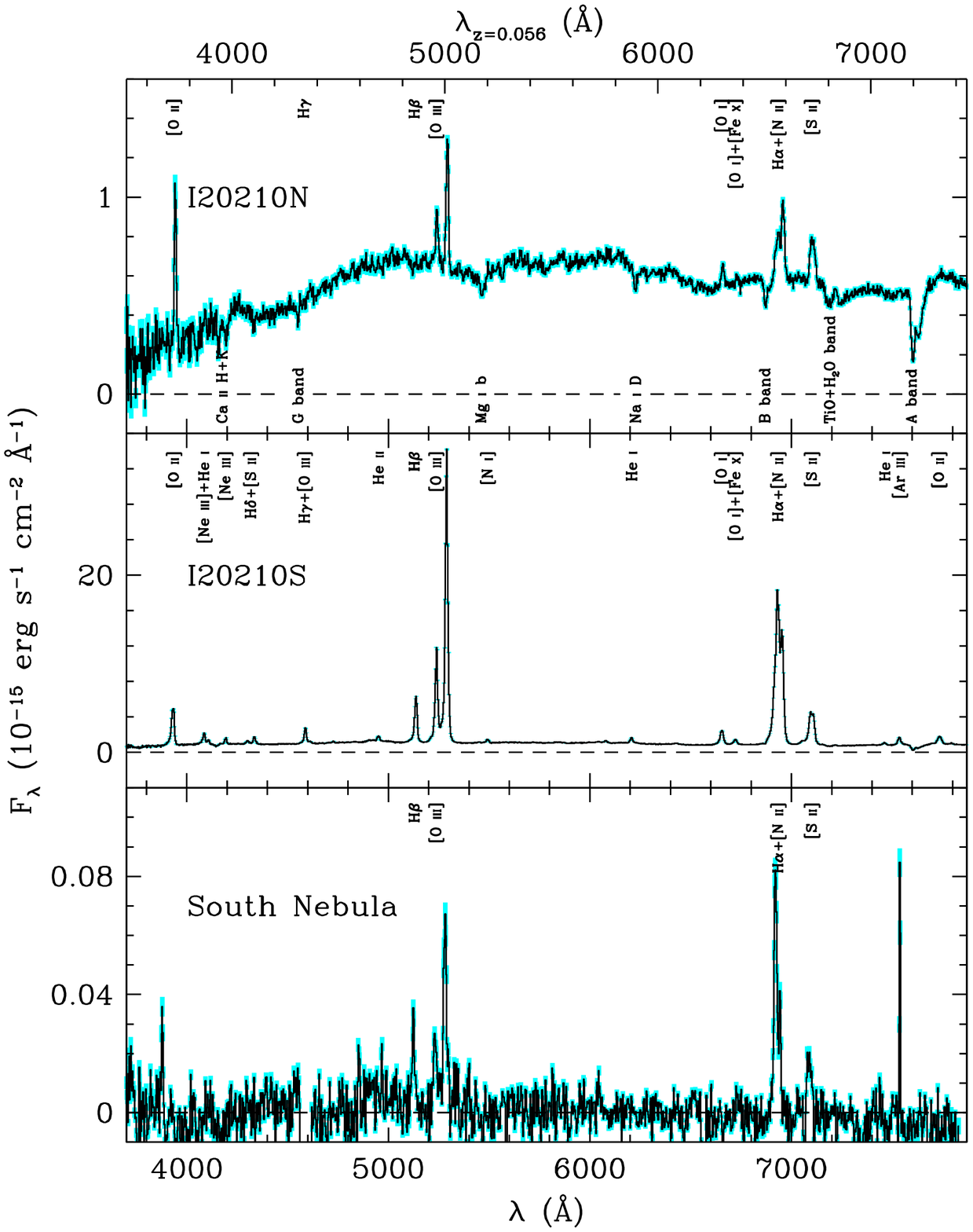}
\caption{Optical spectra of the I20210 components. {\itshape Top panel:} I20210N. {\itshape Middle panel:} I20210S. {\itshape Bottom panel:} The spatially extended South Nebula. In all panels: (i) the detected signal is reported along with its rms uncertainty (cyan bands); (ii) the zero-flux level (dashed line) is indicated; and (iii) the positions of major emission (top) and absorption features (bottom) are labeled accordingly.}
\label{fig:spec}
\end{figure*}

In this work, we present the results of the optical spectroscopic analysis of the $z = 0.056$ dual AGN IRAS 20210+1121 \citep[I20210 hereafter;][P90 hereafter]{Per90}, which is composed of two interacting galaxies oriented in the N-S direction and separated by $12''.2$ \citep[i.e. $\sim$13.3 kpc;][]{Dav02,Arr04}. Considered at first as being composed of a Seyfert 2 with asymmetric emission lines (the southern component) and a normal galaxy (the northern component), X-ray observations performed with XMM-{\itshape Newton} revealed that this system is actually a merger between two obscured AGN hosts \citep{Pic10}, in which the southern member is an ultraluminous infrared galaxy \citep[ULIRG; e.g.,][]{San88}. Additionally, despite having access to spectroscopic data in the near-infrared \citep{Bur01}, the optical spectrum of the northern member was still unobserved, due to its faintness compared to the southern galaxy \citep{Hei95}. The image of the I20210 system, obtained by combining the {\itshape grizy} exposures from the Pan-STARRS1 survey \citep[PS1;][]{Cha16}, is shown in Fig. \ref{fig:photimg}. This image already gives us an idea of the complex structure of the system, displaying a luminous bridge that connects the two galaxies.

This paper is organized as follows. We describe the observation and data-reduction process in Sect. \ref{sec:obsred}. We characterize the extracted spectrum of the northern galaxy in Sect. \ref{sec:emlin}, as well as that of the southern one in Sect. \ref{sec:outneb}. We discuss the relevant physical properties of the structural components of the southern I20210 member in Sects. \ref{sec:disc} and \ref{sec:detneb}. We estimate the supermassive black hole (SMBH) mass of both I20210 members in Sect. \ref{sec:mbhest}. Finally, we summarize our findings in Sect. \ref{sec:conc}. For simplicity, we abbreviate the names of the two galaxies to I20210N (northern member) and I20210S (southern member) hereafter. Throughout the article, we adopt a $\Lambda$-CDM cosmology with \h0, \om\ and \ol.

\section{Observations and data reduction}\label{sec:obsred}

Observations of the I20210 optical spectra were carried out on 2010 August 01 at the {\itshape Telescopio Nazionale Galileo} (TNG; Canarian Islands, Spain). The spectra were simultaneously obtained with the $B$-band grism (wavelength range $\lambda\lambda$3000 -- 8430 \AA, dispersion of 2.52 \AA\ px$^{-1}$, $\lambda/\Delta\lambda = 585$, implying a resolution of 9.8 \AA\ that corresponds to $\sim$510 km s$^{-1}$) of the DOLoRes instrument (point-spread function PSF $\sim 0''.85$), coupled to the $1''.5$ slit. To this end, the instrument configuration was rotated to a position angle of 166$^\circ$ in order to align the slit along the system axis connecting the two nuclei. The two exposures of 600 s each (total exposure time of 1200 s) were then reduced with standard {\footnotesize IRAF} procedures to extract and calibrate the one-dimensional spectra. We show the slit position and orientation (P.A. $= 166^\circ$ east of north) along with the directions of the apertures used to extract the spectra of each object in Fig. \ref{fig:photimg}, superimposed to the PS1 image of the system. The resulting spectra have signal-to-noise ratios of ${\rm S}/{\rm N} \sim 23.6$ (I20210N) and $\sim$33.2 (I20210S), respectively, as computed in line-free continuum regions \citep{Ros12}.

During the extraction and calibration procedures of the 1D spectra, we found that a spectrum emitted from a third location was visible southwards of I20210S, at a projected distance of $\sim$6$''$ (corresponding to $\sim$6.5 kpc given the distance scale of 1.087 kpc/$''$ at $z = 0.056$) from its trace center. This additional spectrum, already identified by P90 in their low-resolution data as extended emission in the I20210S host galaxy (South Nebula, hereafter), is shown in 2D form in Fig. \ref{fig:blobs}: it exhibits the main transitions detected in I20210S (\hb\ $\lambda$4862, \oiii\ $\lambda\lambda$4959,5007, \ha+\n2\ $\lambda\lambda$6548,6583 and \s2\ $\lambda\lambda$6716,6731) detached by $\sim$6$''$ from the I20210S nuclear spectrum, extending over $\sim$2$''$ (i.e. $\sim$2.3 kpc) in the N--S direction and blueshifted by $\sim$450 km s$^{-1}$ with respect to the systemic rest frame. We thus extracted and calibrated it in the same way as we do for the spectra of the main components.

Since the I20210 system is viewed through the Galactic plane, we dereddened the spectra with the Milky Way extinction curve by \citet{Pei92} and $A_{\rm V} = 0.6$ (P90). The final spectra obtained in this way are shown in Fig. \ref{fig:spec}. A visual inspection of the (so-far undetected) I20210N optical spectrum reveals prominent \oii\ $\lambda$3727, \oiii\ $\lambda\lambda$4959,5007, \ha+\n2\ $\lambda\lambda$6548,6583 and \s2\ $\lambda\lambda$6716,6731 emission lines, as well as the lack of the \hb\ $\lambda$4862 feature. Such a spectrum shows strong similarities with those of typical Seyfert 2 galaxies, such as NGC 1667 \citep{Ho93,Ho95,Jon09} and Mrk 1018 \citep{Ost81}.

\section{Characterization of I20210 North}\label{sec:emlin}

We first proceed to estimate the amount of intrinsic dust extinction in each object. To this end, we decided to measure the reddening $E(B-V)$ of the AGN spectrum through the Balmer decrement $F_{{\rm H}\alpha}/F_{{\rm H}\beta}$ \citep[e.g.,][]{Mil72}:
\begin{equation}\label{eqn:baldec}
\frac{F_{{\rm H}\alpha}}{F_{{\rm H}\beta}} = \frac{I_{{\rm H}\alpha}}{I_{{\rm H}\beta}} \cdot 10^{-0.4E(B-V)(1+R_V)(\kappa_\alpha - \kappa_\beta)}
,\end{equation}
with the intrinsic ratio $I_{{\rm H}\alpha}/I_{{\rm H}\beta}$ depending on the physical conditions of the emitting gas only \citep[see e.g.,][and refs. therein]{Gas84}, and with $R_V$ and $\kappa_\lambda$ determined by the adopted extinction model. Since no evidence for narrow lines associated with \hb\ is visible bluewards of the \oiii\ doublet in the I20210N spectrum, we first proceed to model the underlying continuum in order to recover the Balmer emission from the narrow-line region (NLR) of I20210N.

\subsection{I20210N continuum and emission-line fitting}\label{sec:fitprocn}

We modeled the I20210N continuum under the assumption of a negligible AGN contribution to the continuum emission. This is justified by the fact that the I20210N central engine ($L_{\rm X} = 4.7 \times 10^{42}$ erg s$^{-1}$) is highly obscured by a column density $N_{\rm H} \sim 5 \times 10^{23}$ cm$^{-2}$ \citep{Pic10} and, therefore, no light from accretion activity is visible. We subtract the stellar continuum with absorption lines from the I20210N galaxy spectrum by using the penalized-pixel fitting public code {\footnotesize pPXF} \citep{Cap04,Cap12,Cap17}. The spectrum is fitted with a linear combination of stellar spectra templates from the MILES library \citep{Vaz10}, which contains single stellar population synthesis models covering the  same wavelength range as the I20210N spectrum with a full width at half maximum (FWHM) resolution of 2.54 \AA. This procedure also yields information about the kinematics status of the stellar population in the galaxy through the stellar velocity dispersion $\sigma_v^*$.

\begin{figure*}[htbp]
\centering
\includegraphics[scale=0.7,angle=-90]{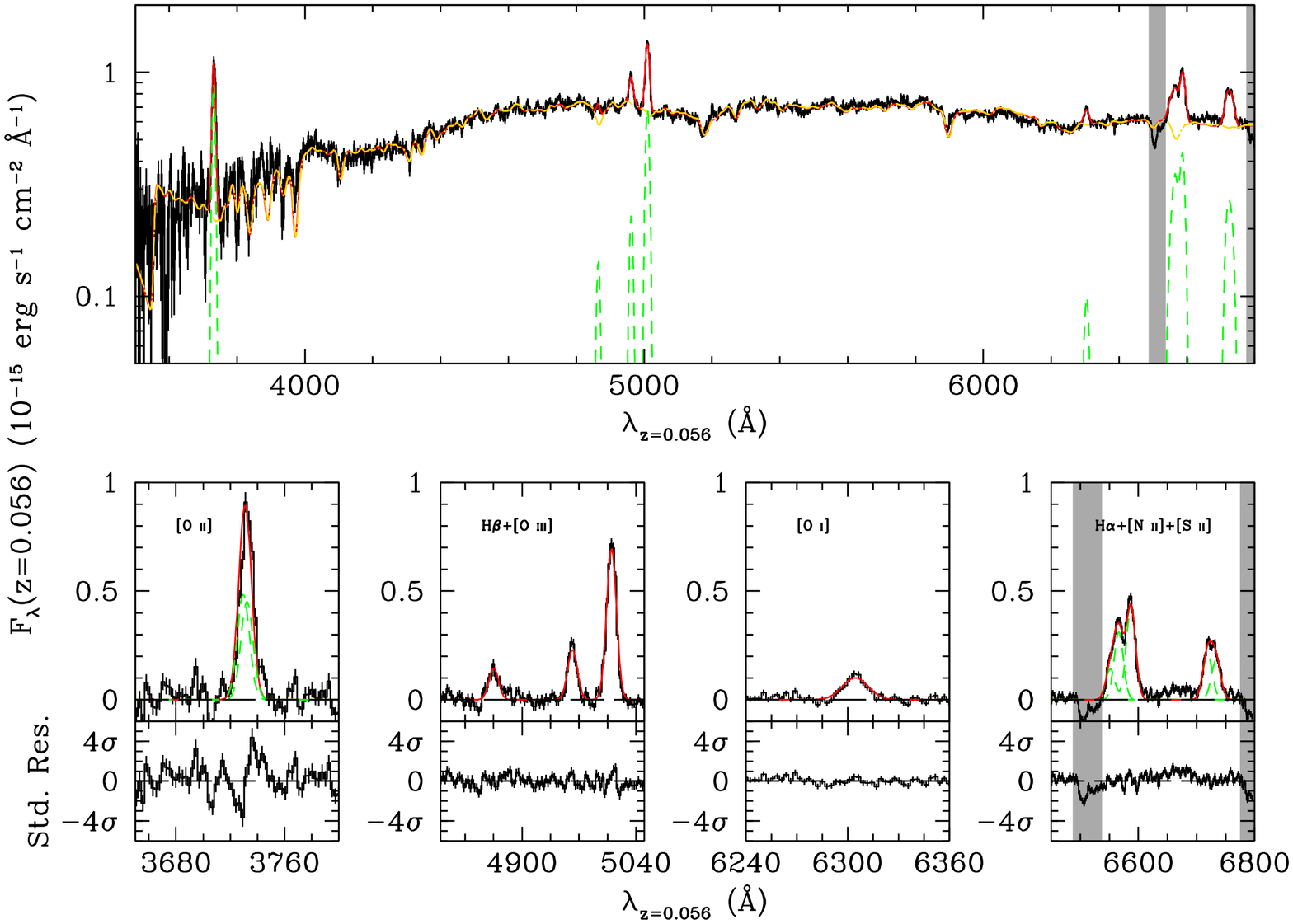}
\caption{Model of the rest-frame spectrum of I20210N. {\itshape Upper panel:} Full I20210N spectrum (black solid line), along with the best-fit starlight model adopted for continuum subtraction (yellow dot-dashed line), the best-fit reddened emission profiles (green short-dashed lines), the global spectral model (red solid line), and the masks applied to the telluric absorption lines (grey bands) shown superimposed to the data. {\itshape Lower panels:} Zoom on the continuum-subtracted emission lines (black solid line), shown along with the global best fit (red solid line) and the best-fit single components (green short-dashed lines) for the blended \oii\ doublet, \ha+\n2\ and \s2\ transitions. The standardized residuals after the best-fit subtraction are also shown in separate windows below each spectral region. In all panels, the zero-level flux (black long-dashed line) is indicated; in the panel with \ha+\n2\ and \s2, the masks applied to the telluric absorption lines (grey bands) are shown superimposed to the data.}
\label{fig:bfitn}
\end{figure*}

We rebinned the MILES templates ($\lambda/\Delta\lambda \sim 2.5$ \AA) to match the DOLoRes spectral resolution of $\sim$10 \AA. We include low-order additive (4th-degree) and multiplicative (1st-degree) Legendre polynomials to adjust the continuum shape of the templates to the observed spectrum. During the fitting procedure, strong emission features are masked out and the spectra are shifted to the rest frame. The {\footnotesize pPXF} best-fit model is chosen through $\chi^2$  minimization. To estimate the uncertainty on the velocity dispersion, we produced $10^3$ realizations of the I20210N spectrum by adding noise to the {\footnotesize pPXF} best-fit model; this noise is drawn from a Gaussian distribution with dispersion equal to the rms of the input spectrum. We then iterate the {\footnotesize pPXF} fitting procedure over such mock spectra and compute the error associated with $\sigma_v^*$ as the standard deviation of the parameter posterior distribution. In doing so, we find a best-fit $\sigma_v^* = 390 \pm 50$ km s$^{-1}$. The residual spectrum obtained by subtracting off the best-fit stellar model from the spectrum is then used to derive emission-line properties. This procedure allowed us to recover the \hb\ narrow emission and therefore compute the Balmer decrement $F_{{\rm H}\alpha}/F_{{\rm H}\beta}$. Both the fitted starlight continuum and the residual emission-line spectrum of I20210N are shown in Fig. \ref{fig:bfitn}.

We note that the derived stellar-velocity dispersion value is very high compared to what is expected for typical galaxies: for example, a search in the catalogue of galactic dynamics by \citet{For99} yields only two elliptical/S0 objects with $\sigma_v^* > 300$ km s$^{-1}$. Similarly, the stellar velocity dispersions measured by \citet{Fal17} in a large sample of galaxies from the CALIFA survey and by \citet{Per21} in a sample of nearby ULIRGs never exceed $\sim$200 km s$^{-1}$. Nevertheless, objects exhibiting exceptional values of $\sigma_v^*$ exist: it is, for instance, the case of NGC 6240 \citep[$\sigma_v^* \sim 360$ km s$^{-1}$;][]{Doy94}, which is indeed a final-state merging system. Therefore, the stellar velocity dispersion value found in I20210N may indicate that the internal kinematics of the galaxy is deeply altered by the gravitational interaction with I20210S.

We then fit the relevant emission lines with Gaussian profiles through the IDL minimization package {\footnotesize MPFIT} \citep{Mar09}. All the narrow components are simultaneously fitted considering them as emitted at the same distance from the AGN, that is, with equal FWHM in the velocity space. In addition, we fix the intensities of the faint components of the \oiii\ and \n2\ doublets to a ratio $1/3.06$ with the respective dominant component \citep[e.g.,][]{Ost06}. In order to compute meaningful uncertainties of measurement for the free parameters, we iterate this process over $10^3$ Monte-Carlo (MC) realizations of each line spectrum. Such realizations have fluxes at each wavelength altered by a random quantity extracted from a Gaussian distribution, which is centered at the specific flux value and wide as the corresponding 1$\sigma$ rms flux error.

The best fit of the I20210N emission lines is shown in Fig. \ref{fig:bfitn}, along with the corresponding standardized residuals\footnote{The standardized residuals are computed everywhere as $\left[F_\lambda - F_\lambda^{\rm (BF)}\right]/\sigma_F$, where $F_\lambda^{\rm (BF)}$ is the best-fit model of the line flux and $\sigma_F$ is the standard deviation of the dimensional residuals $F_\lambda - F_\lambda^{\rm (BF)}$ \citep[e.g.,][]{Coo82}.}. The value of the Balmer decrement derived from this procedure is $2.97 \pm 0.31$, compatible within errors to both the intrinsic ratio $I_{{\rm H}\alpha}/I_{{\rm H}\beta} \sim 2.85$ typical of [H {\scriptsize II}] region-like objects and the AGN ratio of 3.1 \citep{Vei87}. Therefore, we can assume that the NLR of I20210N is viewed along a non-reddened line of sight, with $E(B-V) \sim 0$. The best-fit parameters of the narrow emission lines are reported in Table \ref{tab:pars}, with FWHM corrected for the instrumental broadening $\Delta v_{\rm inst} \sim 510$ km s$^{-1}$ corresponding to the DOLoRes resolution of $\sim$10 \AA:
\begin{equation}\label{eqn:fwhmcorr}
{\rm FWHM}_{\rm corr} = \sqrt{{\rm FWHM}_{\rm obs}^2 - \Delta v_{\rm inst}^2}
.\end{equation}

\begin{sidewaystable*}[htbp]
\centering
\resizebox{\textwidth}{!}{
\begin{tabular}{lcccccccccccc}
\hline
\hline
\multicolumn{13}{l}{ }\\
 & \multicolumn{4}{c}{I20210N ($\chi^2/\nu_{\rm d.o.f.} = 427/405$)} & \multicolumn{5}{c}{I20210S ($\chi^2/\nu_{\rm d.o.f.} = 442/405$)} & \multicolumn{3}{c}{South Nebula ($\chi^2/\nu_{\rm d.o.f.} = 228/209$)}\\
\multicolumn{13}{l}{ }\\
\cline{2-13}
\multicolumn{13}{l}{ }\\
Transition & \multicolumn{2}{c}{Flux ($10^{-14}$ erg s$^{-1}$ cm$^{-2}$)} & \multicolumn{2}{c}{FWHM (km s$^{-1}$)} & \multicolumn{2}{c}{Flux ($10^{-14}$ erg s$^{-1}$ cm$^{-2}$)} & \multicolumn{2}{c}{FWHM (km s$^{-1}$)} & Blueshift (km s$^{-1}$) & Flux ($10^{-14}$ erg s$^{-1}$ cm$^{-2}$) & FWHM (km s$^{-1}$) & Blueshift (km s$^{-1}$)\\
\multicolumn{13}{l}{ }\\
\cline{6-10}
\multicolumn{13}{l}{ }\\
\multicolumn{5}{l}{ } & N & B & N & B & |N--B|\\
\multicolumn{13}{l}{ }\\
\hline
\multicolumn{13}{l}{ }\\
\oii\ $\lambda\lambda$3726,3729 & \multicolumn{2}{c}{$1.060 \pm 0.055$} &  \multicolumn{2}{c}{$690 \pm 40$} & \multicolumn{1}{c}{$10.4 \pm 1.1$} & \multicolumn{1}{c}{$10.6 \pm 4.4$} & \multicolumn{1}{c}{$530 \pm 90$} & \multicolumn{1}{c}{*} & * & --- & --- & ---\\
\oiii\ $\lambda$4363 & \multicolumn{2}{c}{---} & \multicolumn{2}{c}{---} & \multicolumn{1}{c}{$0.500 \pm 0.054$} & \multicolumn{1}{c}{$0.42 \pm 0.17$} & \multicolumn{1}{c}{''} & \multicolumn{1}{c}{*} & * & --- & --- & ---\\
\hb\ & \multicolumn{2}{c}{$0.214 \pm 0.010$} & \multicolumn{2}{c}{''} & \multicolumn{1}{c}{$13.72 \pm 0.94$} & \multicolumn{1}{c}{$6.3 \pm 2.1$} & \multicolumn{1}{c}{''} & \multicolumn{1}{c}{$1960 \pm 470$} & $330 \pm 180$ & $0.0691 \pm 0.0036$ & $710 \pm 330$ & $550 \pm 150$\\
\oiii\ $\lambda$5007 & \multicolumn{2}{c}{$1.058 \pm 0.047$} & \multicolumn{2}{c}{''} & \multicolumn{1}{c}{$85.2 \pm 5.4$} & \multicolumn{1}{c}{$32.1 \pm 9.8$} & \multicolumn{1}{c}{''} & \multicolumn{1}{c}{$2080 \pm 310$} & $390 \pm 120$ & $0.1193 \pm 0.0073$ & '' & ''\\
\oi\ $\lambda$6300 & \multicolumn{2}{c}{$0.193 \pm 0.010$} & \multicolumn{2}{c}{''} & \multicolumn{1}{c}{$4.13 \pm 0.20$} & \multicolumn{1}{c}{$2.39 \pm 0.57$} & \multicolumn{1}{c}{''} & \multicolumn{1}{c}{$1910 \pm 350$} & $230 \pm 150$ & --- & --- & ---\\
\ha\ & \multicolumn{2}{c}{$0.623 \pm 0.028$} & \multicolumn{2}{c}{''} & \multicolumn{1}{c}{$42.5 \pm 2.0$} & \multicolumn{1}{c}{$19.5 \pm 4.3$} & \multicolumn{1}{c}{''} & \multicolumn{1}{c}{*}  & * & $0.1418 \pm 0.0071$ & '' & ''\\
\n2\ $\lambda$6583 & \multicolumn{2}{c}{$0.863 \pm 0.039$} & \multicolumn{2}{c}{''} & \multicolumn{1}{c}{$33.7 \pm 1.6$} & \multicolumn{1}{c}{$8.0 \pm 2.2$} & \multicolumn{1}{c}{''} & \multicolumn{1}{c}{*}  & * & $0.0334 \pm 0.0047$ & '' & ''\\
\s2\ $\lambda$6717 & \multicolumn{2}{c}{$0.436 \pm 0.021$} & \multicolumn{2}{c}{''} & \multicolumn{1}{c}{$6.80 \pm 0.31$} & \multicolumn{1}{c}{$2.98 \pm 0.66$} & \multicolumn{1}{c}{''} & \multicolumn{1}{c}{*}  & * & $0.0371 \pm 0.0012$ & '' & ''\\
\s2\ $\lambda$6731 & \multicolumn{2}{c}{$0.366 \pm 0.018$} & \multicolumn{2}{c}{''} & \multicolumn{1}{c}{$7.21 \pm 0.32$} & \multicolumn{1}{c}{$5.0 \pm 1.0$} & \multicolumn{1}{c}{''} & \multicolumn{1}{c}{*}  & * & $0.0273 \pm 0.0010$ & '' & ''\\
\multicolumn{13}{l}{ }\\
\hline
\multicolumn{13}{l}{ }\\
$E(B-V)$ & \multicolumn{2}{c}{$\sim$0} & \multicolumn{2}{c}{ } & $0.271 \pm 0.019$ & $0.195 \pm 0.091$ & & & & $\sim$0 & \\
\multicolumn{13}{l}{ }\\
\hline
\hline
\multicolumn{13}{l}{ }\\
\multicolumn{1}{c}{ } & \multicolumn{3}{c}{I20210N} & \multicolumn{3}{c}{I20210S} & \multicolumn{3}{c}{Outflow} & \multicolumn{3}{c}{South Nebula}\\
\multicolumn{13}{l}{ }\\
\cline{2-13}
\multicolumn{13}{l}{ }\\
Ratio & \multicolumn{2}{c}{Value} & \multicolumn{1}{c}{Uncertainty} & \multicolumn{2}{c}{Value} & \multicolumn{1}{c}{Uncertainty} & \multicolumn{2}{c}{Value} & \multicolumn{1}{c}{Uncertainty} & \multicolumn{2}{c}{Value} & \multicolumn{1}{c}{Uncertainty}\\
\multicolumn{13}{l}{ }\\
\hline
\multicolumn{13}{l}{ }\\
$\log{\left(
\mbox{\oiii}/\mbox{\hb}
\right)}$ & \multicolumn{2}{c}{$0.686$} & \multicolumn{1}{c}{$0.040$} &  \multicolumn{2}{c}{$0.798$} & \multicolumn{1}{c}{$0.061$} & \multicolumn{2}{c}{$0.71$} & \multicolumn{1}{c}{$0.28$} & \multicolumn{2}{c}{$0.237$} & \multicolumn{1}{c}{$0.049$}\\
$\log{\left(
\mbox{\n2}/\mbox{\ha}
\right)}$ & \multicolumn{2}{c}{$0.142$} & \multicolumn{1}{c}{$0.039$} &  \multicolumn{2}{c}{$-0.114$} & \multicolumn{1}{c}{$0.044$} & \multicolumn{2}{c}{$-0.39$} & \multicolumn{1}{c}{$0.21$} & \multicolumn{2}{c}{$-0.628$} & \multicolumn{1}{c}{$0.083$}\\
$\log{\left(
\mbox{\s2}/\mbox{\ha}
\right)}$ & \multicolumn{2}{c}{$0.110$} & \multicolumn{1}{c}{$0.041$} & \multicolumn{2}{c}{$-0.489$} & \multicolumn{1}{c}{$0.044$} & \multicolumn{2}{c}{$-0.39$} & \multicolumn{1}{c}{$0.19$} & \multicolumn{2}{c}{$-0.343$} & \multicolumn{1}{c}{$0.037$}\\
$\log{\left(
\mbox{\oi}/\mbox{\ha}
\right)}$ & \multicolumn{2}{c}{$-0.509$} & \multicolumn{1}{c}{$0.042$} &  \multicolumn{2}{c}{$-0.933$} & \multicolumn{1}{c}{$0.052$} & \multicolumn{2}{c}{$-0.91$} & \multicolumn{1}{c}{$0.20$} & \multicolumn{2}{c}{---} & \multicolumn{1}{c}{---}\\
$\log{\left(
\mbox{\oii}/\mbox{\hb}
\right)}$ & \multicolumn{2}{c}{$0.695$} & \multicolumn{1}{c}{$0.043$} & \multicolumn{2}{c}{$-0.120$} & \multicolumn{1}{c}{$0.077$} & \multicolumn{2}{c}{$0.23$} & \multicolumn{1}{c}{$0.33$} & \multicolumn{2}{c}{---} & \multicolumn{1}{c}{---}\\
$\log{\left(
\mbox{\oiii}/\mbox{\oii}
\right)}$ & \multicolumn{2}{c}{$-0.001$} & \multicolumn{1}{c}{$0.042$} &  \multicolumn{2}{c}{$0.913$} & \multicolumn{1}{c}{$0.073$} & \multicolumn{2}{c}{$0.48$} & \multicolumn{1}{c}{$0.31$} & \multicolumn{2}{c}{---} & \multicolumn{1}{c}{---}\\
\multicolumn{13}{l}{ }\\
\hline
\multicolumn{13}{l}{$^-$Transition not detected in the spectrum.}\\
\multicolumn{13}{l}{$^{{\rm ''}}$Value fixed to be equal to the first non-null one upwards in the column.}\\
\multicolumn{13}{l}{$^*$Value anchored to the best fit of the \oiii\ $\lambda$5007 broad component.}\\
\end{tabular}
}
\caption{{\itshape Top}: Best-fit parameters of the dereddened diagnostic emission lines in the optical spectrum of I20210N, I20210S NLR (``N'' components) and outflowing emission (``B'' components), and emission from the South Nebula. The integrated fluxes presented here have been dereddened by the indicated amount of $E(B-V)$ with the SMC extinction by \citet{Pei92}. {\itshape Bottom}: Values of the diagnostic line ratios with the corresponding uncertainties.}
\label{tab:pars}
\end{sidewaystable*}

\begin{figure*}[htbp]
\centering
\includegraphics[scale=0.8,angle=90]{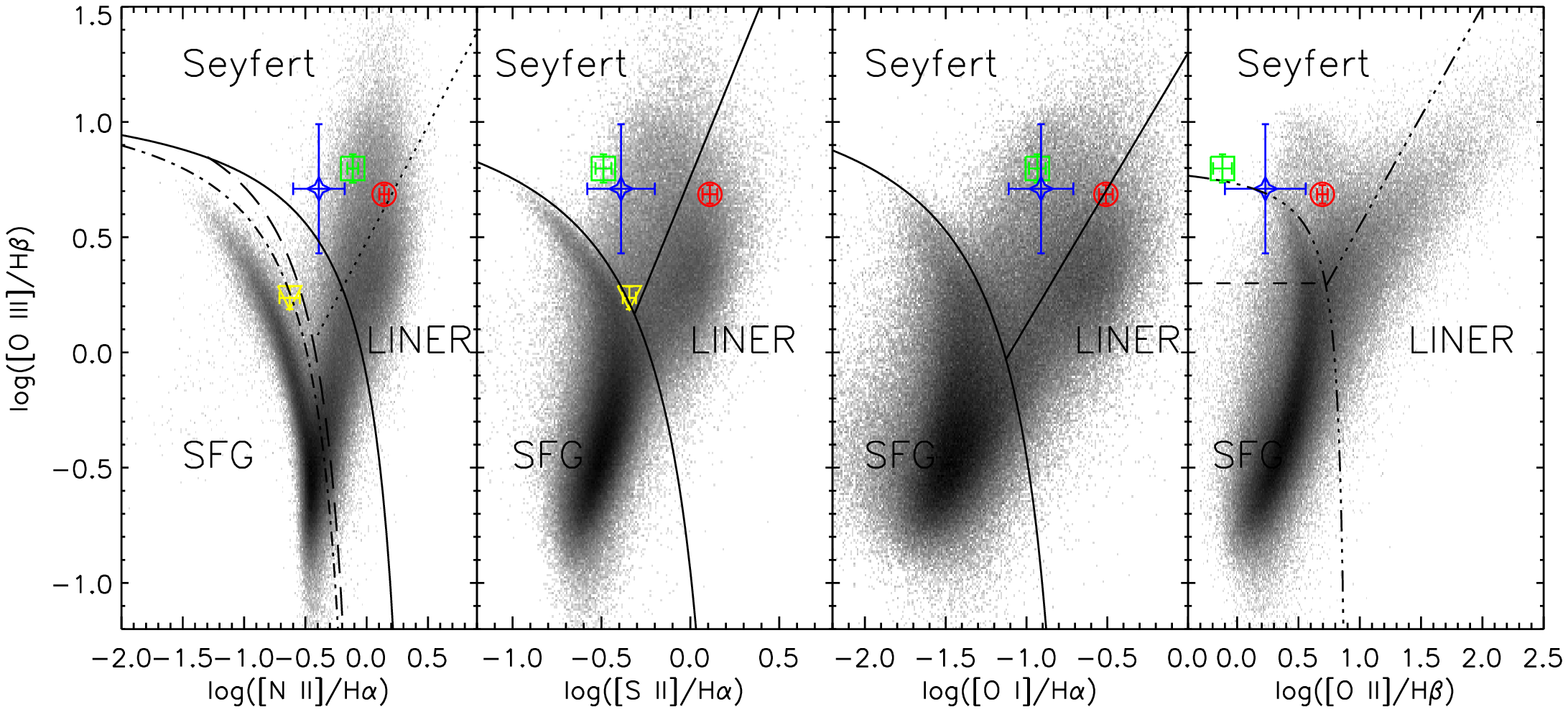}
\caption{BPT diagnostic diagrams showing the position of I20210N (red circle), I20210S (green square) and its outflow (blue star), and the South Nebula (yellow triangle), along with the relative uncertainties on top of the SDSS data from the OSSY database \citep[][grey dots]{Oh15}. As a reference, in the first three panels, the extreme-starburst and Seyfert-LINER classification boundaries by \citet[][solid lines]{Kew01} are indicated, along with the pure star-formation boundary by \citet[][long-dashed line]{Kau03}, the alternative Seyfert-LINER relation by \citet[][dotted line]{Cid10}, and the redshift-dependent relation at $z \sim 0.13$ by \citet[][dot-dashed line]{Kew13a} in the \oiii/\hb-to-\n2/\ha\ diagram. In the \oiii/\hb-to-\oii/\hb\ diagram, the star-forming and Seyfert-LINER boundaries by \citet[triple dot-dashed line]{Lam10} are indicated, along with the mixed-region boundary (short-dashed line).}
\label{fig:bpt}
\end{figure*}

\subsection{I20210N classification}\label{sec:classn}

To assess the nature of the AGN hosted in I20210N, we computed the \oiii/\hb, \n2/\ha, \s2/\ha, \oi/\ha\ and \oii/\hb\ logarithmic line ratios from the best-fit emission line parameters. The derived values are presented in Table \ref{tab:pars}, along with their uncertainties. For completeness, we also report the value of the \hbox{\oiii/\oii} ratio that is used to further classify active galaxies \citep[e.g.,][]{Hec80,Kew06}.

The values derived for I20210N are plotted in the BPT diagrams shown in Fig. \ref{fig:bpt}, superimposed to the values for SDSS-DR7 objects retrieved from the OSSY database \citep{Oh11,Oh15}. To discriminate between the different classes of emission-line galaxies (star-forming, Seyferts, LINERs), we adopt from the current literature the relations defining the boundaries between types of galactic activity: (i) the extreme-starburst relation and the Seyfert-LINER boundaries by \citet{Kew01} in the original diagrams by \citet{Bal81}; (ii) the star-forming, Seyfert-LINER and mixed-region boundaries by \citet{Lam10} in the \oiii/\hb-to-\oii/\hb\ diagram; (iii) for the \oiii/\hb-to-\n2/\ha\ diagram, the pure star-formation boundary by \citet{Kau03}, the alternative Seyfert-LINER relation by \citet{Cid10} and the redshift-dependent star formation boundary by \citet[][see also \citealt{Kew13b}]{Kew13a} computed for $z = 0.128 \pm 0.044$, that is, the average redshift value of the OSSY catalog \citep{Oh15}.

Figure \ref{fig:bpt} shows that the I20210N line ratios are generally consistent with those found for Seyfert galaxies in the \oiii/\hb-to-\n2/\ha\ and \oiii/\hb-to-\oii/\hb\ diagrams, while they remain intermediate between a Seyfert and a LINER in the \oiii/\hb-to-\s2/\ha\ diagram, where they sit on the Seyfert-to-LINER boundary by \citet{Kew01}. Due to their intermediate nature, \citet{Hec80} calls these kinds of objects ``transition galaxies'' that lie between Seyfert 1.9 \citep{Ost81} and LINERs. However, the severe unresolved blending that affects the \ha+\n2\ emission system may offer an alternative explanation: in fact, decreasing intrinsic \ha\ intensity would eventually move the position of I20210N in both the \oiii/\hb-to-\s2/\ha\ and \oiii/\hb-to-\oi/\ha\ diagrams in the Seyfert region (neglecting the shift in the \oiii/\hb-to-\n2/\ha\ diagram given by the consequent increase in the \n2\ emission). In addition, the hard X-ray luminosity of $\sim$5$\times$10$^{42}$ erg s$^{-1}$ measured for I20210N by \citet{Pic10} helps in breaking this uncertainty, pushing the classification of I20210N towards a Seyfert 2 galaxy.

\begin{figure*}[htbp]
\begin{center}
\includegraphics[scale=0.7,angle=-90]{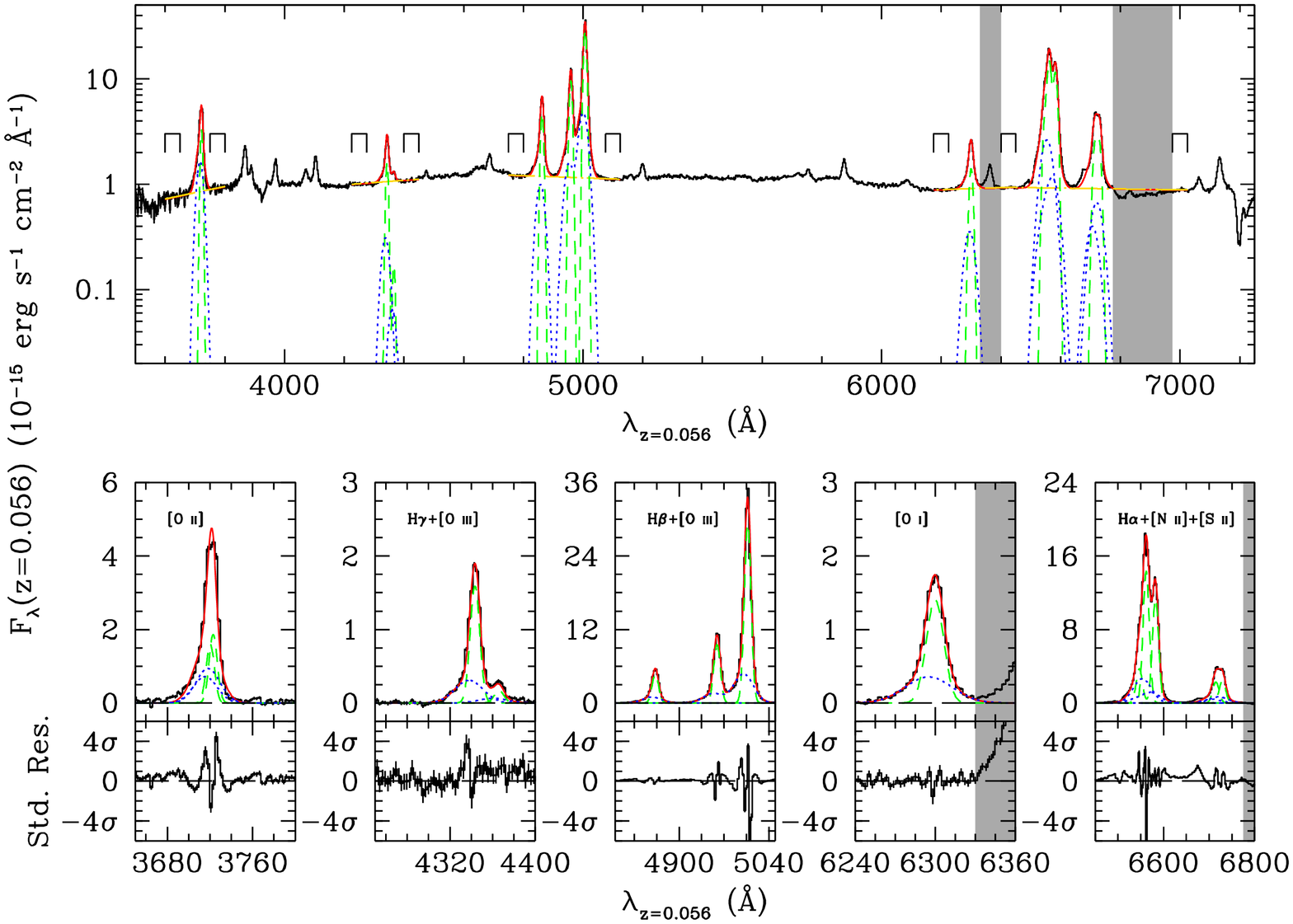}
\end{center}
\caption{Model of the rest-frame spectrum of I20210S. {\itshape Upper panel:} Full I20210S spectrum (black solid line), along with the local power laws adopted for continuum subtraction (yellow dot-dashed lines), the best-fit reddened narrow (green short-dashed lines) and broad emission profiles (blue dotted lines), and the global spectral model (red solid line) shown superimposed to the data. The intervals used for the local continuum subtraction under the emission lines are marked, along with the masks applied to the \oi+[Fe {\scriptsize X}] emission and to the telluric H$_2$O absorption line (grey bands). {\itshape Lower panel:} Zoom on the continuum-subtracted emission lines (black solid line), shown along with the global best fit (red solid line) and the best-fit narrow (green short-dashed lines) and broad components (blue dotted lines). The standardized residuals after the best-fit subtraction are also shown in separate windows below each spectral region. In all panels, the zero-level flux (black long-dashed line) is indicated; in the panel with \oi, the mask applied to the \oi+[Fe {\scriptsize X}] emission line is shown superimposed to the data, as well as the one for the H$_2$O telluric absorption in the panel with \ha+\n2 and \s2 (grey bands).}
\label{fig:bfits}
\end{figure*}

\section{Characterization of I20210 South}\label{sec:outneb}

For both the nuclear spectrum and the South Nebula, we applied the same fitting procedure as done for I20210N. Therefore, we first proceed to model the I20210S continuum emission. We assume for I20210S an AGN-dominated object, that is, one exhibiting a power-law continuum reddened by foreground dust; such an assumption is well motivated by the quasar-like energetics of I20210S \citep[e.g., $L_{\rm bol} \sim 10^{45}$ erg s$^{-1}$;][]{Pic10}.

\subsection{I20210S nuclear continuum and emission-line fitting}\label{sec:fitprocs}

The optical spectrum of I20210S is heavily reddened by intrinsic dust, which is particularly evident by the lack of a flux rise bluewards the \oii\ emission. However, due to the presence of a large number of emission and telluric lines that greatly reduce the intervals of featureless spectral regions, we decided not to perform a global continuum fit to be subtracted from the spectrum. Instead, we selected sections of the I20210S spectrum free of major features and adjacent to the lines of interest, and interpolated them with local power laws in order to remove the underlying AGN emission (see Fig. \ref{fig:bfits}).

Then we used the MC fitting procedure to model the emission lines with Gaussian profiles, which are shown in Fig. \ref{fig:bfits}, along with the corresponding standardized residuals. Differently from I20210N, we used two components for each transition to account for the total emission profile in the spectrum, as was also done by \citet{Arr14}. The main narrow components obey the prescriptions presented in Sect. \ref{sec:fitprocn}; the additional emission, as already found by P90, consists in a broad line blueshifted by $\sim$400 km s$^{-1}$ with FWHM $\sim 2000$ km s$^{-1}$. The possibility that such features are an effect of the orientation of the I20210S narrow-line region (NLR) as described in \citet{Bio17} is ruled out. In fact, for emission lines associated with permitted transitions the asymmetries would be redshifted with respect to the line center (see e.g., their figure 3). Therefore, we can conclude that I20210S exhibits evidence of an ionized gas outflow.

\begin{figure*}[htbp]
\centering
\begin{minipage}{.49\textwidth}
\includegraphics[width=\textwidth]{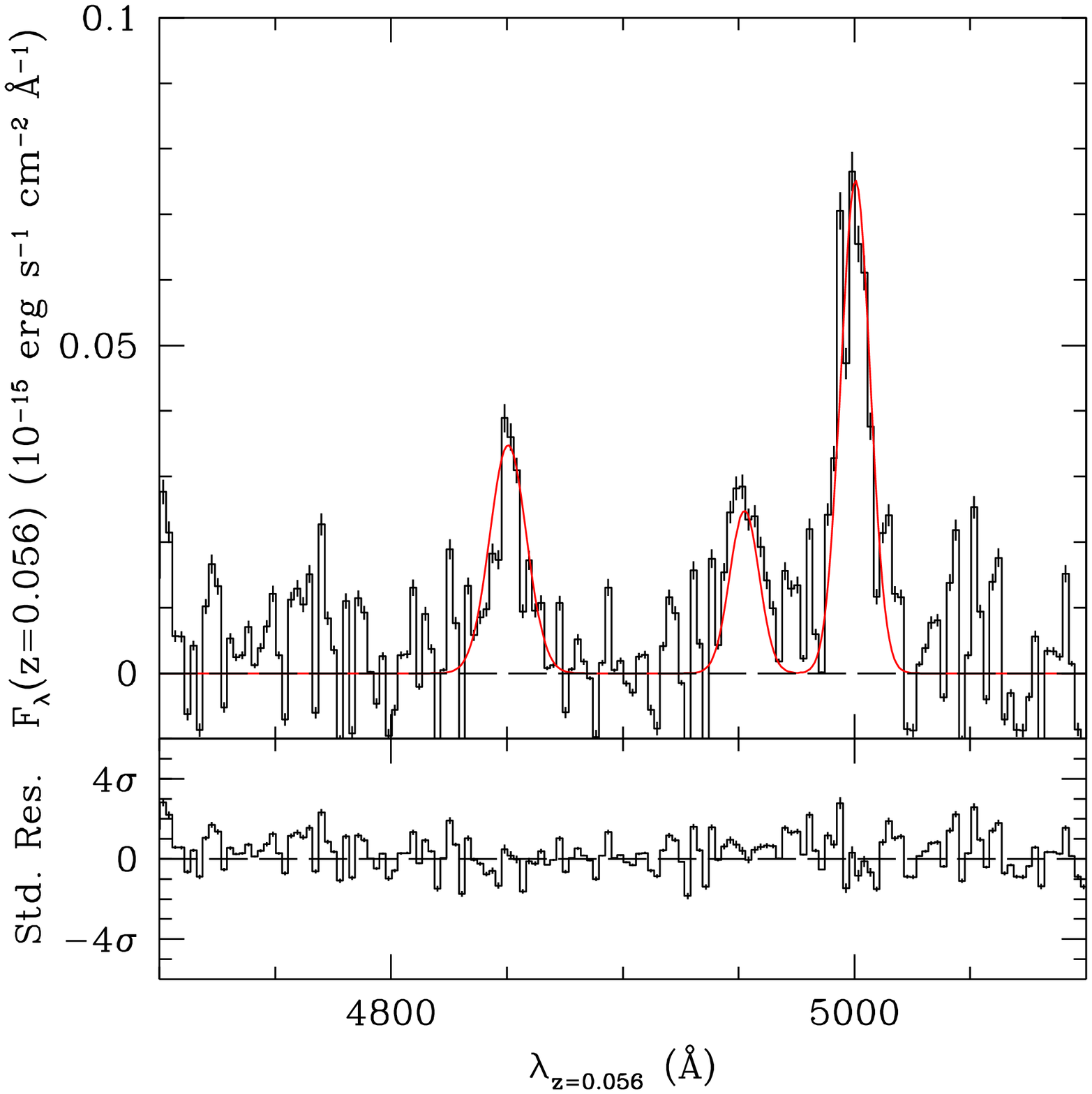}
\end{minipage}
\begin{minipage}{.49\textwidth}
\includegraphics[width=\textwidth]{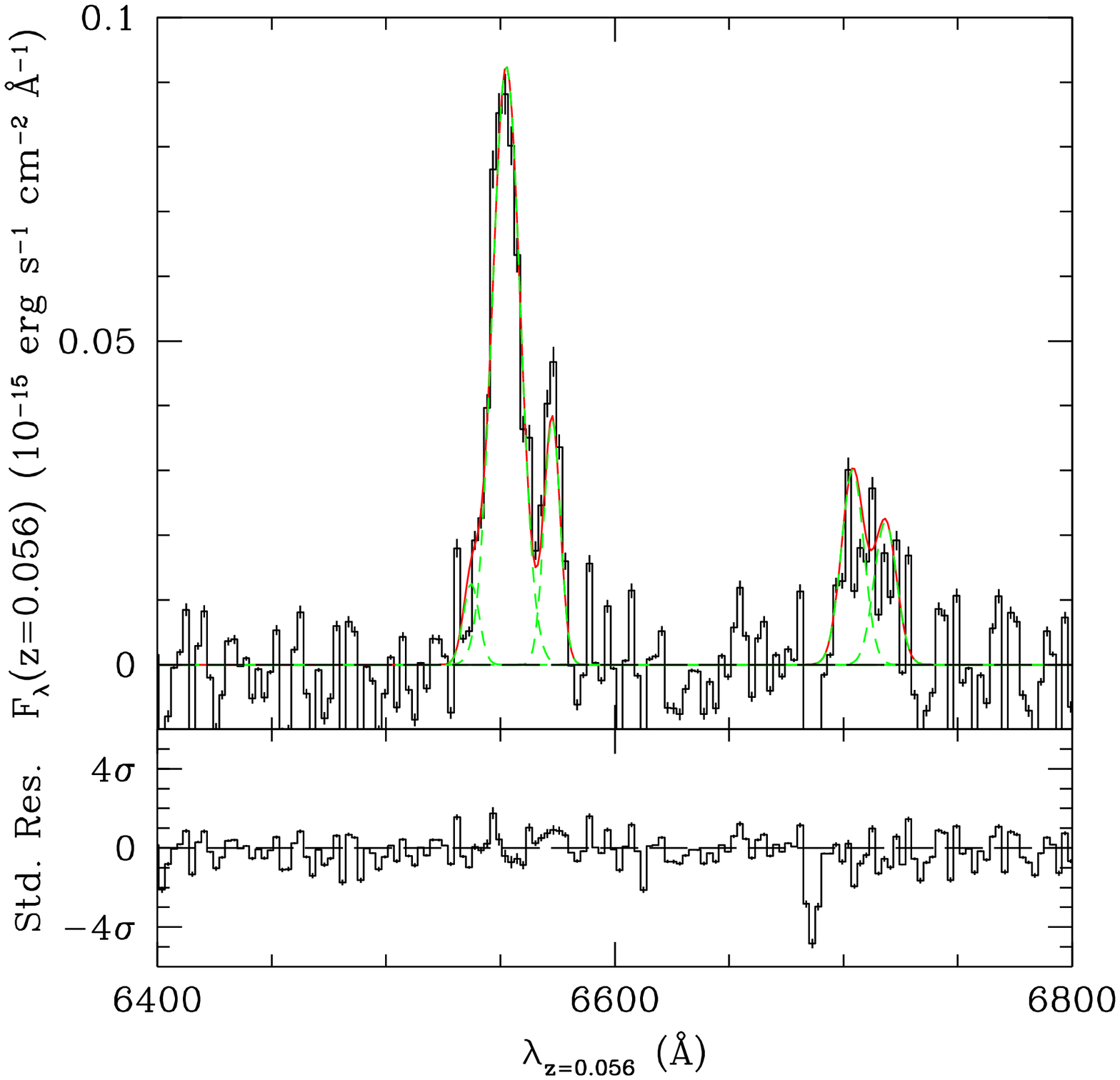}
\end{minipage}
\caption{Best-fit profiles of the emission lines from the South Nebula in the rest frame. {\itshape Left panel:} The \hb+\oiii\ spectral region. {\itshape Right panel:} The \ha+\n2\ and \s2\ spectral region. In both panels, the global emission profile (red solid line) is shown superimposed to the line spectrum (black histogram) along with the single components of the \ha+\n2\ and \s2\ blended profiles (green short-dashed lines), and the zero-flux level (black long-dashed line) is indicated. The standardized residuals after the best-fit subtraction are also shown in separate windows below each spectral region.}
\label{fig:nebfit}
\end{figure*}

To account for this additional emission, we include the broad components in the fit of the I20210S line profiles anchoring the blueshift and FWHM values of the transitions affected by severe blending -- namely, the \oii\ $\lambda\lambda$3726,3729 doublet, the H$\gamma$+\oiii\ $\lambda$4363, the \ha+\n2\ system and the \s2\ doublet -- to those of the \oiii\ $\lambda$5007 (see Table \ref{tab:pars}). This choice is motivated by the fact that the \oiii\ emission has the highest S/N; in the cases where an anchoring to its parameters is adopted, only the line amplitude is left free to vary. In addition, we estimate the amount of intrinsic dust extinction for the NLR and the outflow separately since the two regions are, in principle, located at different distances from the central engine and can thus be affected by different amounts of reddening.

The Balmer ratio derived from the MC fit for the narrow components is $F_{{\rm H}\alpha} / F_{{\rm H}\beta} = 4.108 \pm 0.079$, corresponding to $E(B-V)_{\rm NLR} = 0.271 \pm 0.019$ mag for the AGN intrinsic ratio $I_{{\rm H}\alpha}/I_{{\rm H}\beta} = 3.1$ \citep{Vei87} and the SMC extinction by \citet{Pei92}, whereas a ratio $F_{{\rm H}\alpha} / F_{{\rm H}\beta} = 3.80 \pm 0.34$ for the outflow yields $E(B-V)_{\rm out} = 0.195 \pm 0.091$. Finally, we applied the extinctions derived in this way to deredden the corresponding emission-line amplitudes. The best fit of the reddened I20210S spectrum is shown in Fig. \ref{fig:bfits}, whereas the best-fit parameters of both its NLR and outflow emission are reported in Table \ref{tab:pars}. On average, the I20210S wind has an outflow velocity $\Delta v = 330 \pm 170$ km s$^{-1}$ and ${\rm FWHM} = 2000 \pm 390$ km s$^{-1}$: such values are a factor of $\sim$2.4 higher than the corresponding mean parameters found by \citet{Arr14} in ULIRGs hosting AGN (see their table 2), and more in line with those found by \citet{Rod13} for ionized outflows in nearby ULIRGs (see their Table 2) and by \citet{Zak16} in high-$z$ reddened quasars (see their table 1) where the emission-line profiles are modeled using multiple Gaussians.

\subsection{South Nebula spectrum}\label{sec:nebula}

Next we estimated the intrinsic reddening of the South Nebula. The detection of both \hb\ and \ha\ narrow transitions allows us to apply the MC line-fitting procedure described in the case of I20210N (see Fig. \ref{fig:nebfit} and Sect. \ref{sec:fitprocn}). Lacking any trace of an underlying continuum that could have been used in the determination of the reddening law, we adopted the SMC extinction by \citet{Pei92}, as in the case of the parent nucleus. This in turn yields $F_{{\rm H}\alpha}/F_{{\rm H}\beta} = 2.05 \pm 0.10$, which is lower than the intrinsic ratio of 2.85 valid for [H {\scriptsize II}] regions. Also, the low associated error of measurement potentially indicates a poorly determined estimate of the Balmer ratio, likely due to the uncertainties in extracting a continuum-less spectrum that is $\sim$100 times less intense than the I20210S nuclear emission. Therefore, the issue of determining the South Nebula intrinsic reddening is clearly a matter that ought to be left to more sensitive, spatially resolved spectroscopic future data; in the following, we consider it compatible with $E(B-V) \sim 0$.

We then fit the parameters of the five emission features that are clearly identified, namely \hb, \oiii, \ha, \n2\ and \s2, without applying any dereddening (see Table \ref{tab:pars}). Interestingly, the South Nebula exhibits a blueshift of $550 \pm 150$ km s$^{-1}$ with respect to the systemic redshift and a FWHM of $710 \pm 330$ km s$^{-1}$. Such features are a clear indication of highly disrupted gas \citep{Bel13}, similar to that found by \citet{Ram17} in the Teacup Galaxy \citep[$L_{\rm [OIII]} \sim 5 \times 10^{42}$ erg s$^{-1}$ according to][to be compared with $L_{\rm [OIII]} \sim 6.5 \times 10^{42}$ erg s$^{-1}$ for I20210S]{Rey08} at comparable distances ($\sim$5.6 kpc) from the central engine.

\subsection{I20210S classification}\label{sec:classs}

As done in Sect. \ref{sec:classn} for I20210N, we computed the line ratios for all the regions decomposed from the spectrum of I20210S, namely, the NLR, the outflow and the South Nebula, and we placed them in the relevant BPT diagrams to obtain a first discrimination between an AGN or star-formation powered emission. A visual inspection confirms that the NLR properties are fully consistent with their AGN nature, as well as with the outflow emission falling well inside the AGN region shown to be in agreement with the scenario of an ionized wind driven by the nuclear activity.

The South Nebula sits close enough to the boundary between AGN and star-forming galaxies to prevent its straightforward inclusion among the AGN-powered processes. However, the kinematic properties of this region (velocity blueshift of $\sim$500 km s$^{-1}$, FWHM of $\sim$700 km s$^{-1}$) may  actually be interpreted as being due to the I20210S outflow, which has stripped out ionized gas from the I20210S nucleus. The possibility that the South Nebula is an extended NLR component blown out of the central engine by radiation pressure is in principle supported by studies that ubiquitously find NLRs extended over $\sim$10 kpc from the central engine in both Type 1 and Type 2 quasars \citep[see e.g.][and refs. therein]{Hus13}; in the case of I20210S, this is less likely and it is expected, rather, to be extended, non-outflowing gas associated with extreme mergers that is due to the fact that blown-out NLRs typically show ${\rm FWHM} < 250$ km s$^{-1}$ \citep{Bel13}.

Our data do not allow a deeper exploration of the spectral properties of the South Nebula. Therefore, we point out that due to the intermediate values of its diagnostic line ratios between AGN and star-forming galaxies, this detached emitting region is a very interesting environment in which the effects of AGN feedback may be at work in pushing the gas outside the central region of the host galaxy (negative feedback) while also triggering some amount of star formation into it \citep[positive feedback; see e.g.,][]{Mai17}. Thus, it is worthy of further investigation with high-quality spatially resolved spectroscopy.

\section{Physical properties of the outflow in I20210S}\label{sec:disc}

Next we were able to characterize the physics of the I20120S ionized wind. To this aim, we estimated the outflowing mass $M_{\rm out}$ and the mass loss rate $\dot{M}$ of the ionized gas following the method presented in \citet{Kak16} and \citet{Bis17}. Under the assumptions that (i) the AGN wind is free, spherically or biconically symmetric, and mass-conserving; (ii) the AGN wind has mass-outflow rate and velocity independent on the outflow radius \citep{Rup02,Rup05}; and (iii) most of the oxygen consists of O$^{2+}$ ions, we can use the relation by \citet{Car15}:
\begin{equation}\label{eqn:mout}
\resizebox{0.9\hsize}{!}{$
\log{\left(
\frac{M_{\rm out}}{{\rm M}_\odot}
\right)} = 7.6 + \log{\left(
\frac{C}{10^{{\rm [O/H]} - {\rm [O/H]}_\odot}}
\right)} + \log{\left(
\frac{L^{\rm out}_{\rm [OIII]}}{10^{44}\mbox{ erg s}^{-1}}
\right)} - \log{\left(
\frac{\langle n_e \rangle}{10^3\mbox{ cm}^{-3}}
\right)}
$}
\end{equation}
where $C = \langle n_e \rangle^2 / \langle n_e^2 \rangle$, ${\rm [O/H]} - {\rm [O/H]}_\odot$ is the gas metallicity relative to the solar value, $L^{\rm out}_{\rm [OIII]}$ is the outflowing \oiii\ $\lambda 5007$ luminosity, and $\langle n_e \rangle$ is the average electron density. The latter is, in turn, related to the electron temperature, $T_e$, which can be derived from the line ratios $\left( I_{4959} + I_{5007} \right)/I_{4363}$ of the outflow emission \citep{Ost06}:
\begin{equation}\label{eqn:temp}
\frac{I_{4959} + I_{5007}}{I_{4363}} \approx 7.90 \cdot \exp{\left(
\frac{32,900\mbox{ K}}{T_e}
\right)}
.\end{equation}
From the decomposition of the emission lines in the spectrum of I20210S through the fit with multiple Gaussian components described in Sect. \ref{sec:emlin} (see Fig. \ref{fig:bfits}), we compute $\left( I_{4959} + I_{5007} \right)/I_{4363} = 100 \pm 70$, which corresponds to $T_e = 12,900 \pm 3600$ K and is in agreement with the value of $\sim$10$^4$ K generally assumed for AGN outflows \citep[see e g.][and refs. therein]{Per17b}.

\begin{figure*}[htbp]
\centering
\begin{minipage}{.49\textwidth}
\includegraphics[width=\textwidth]{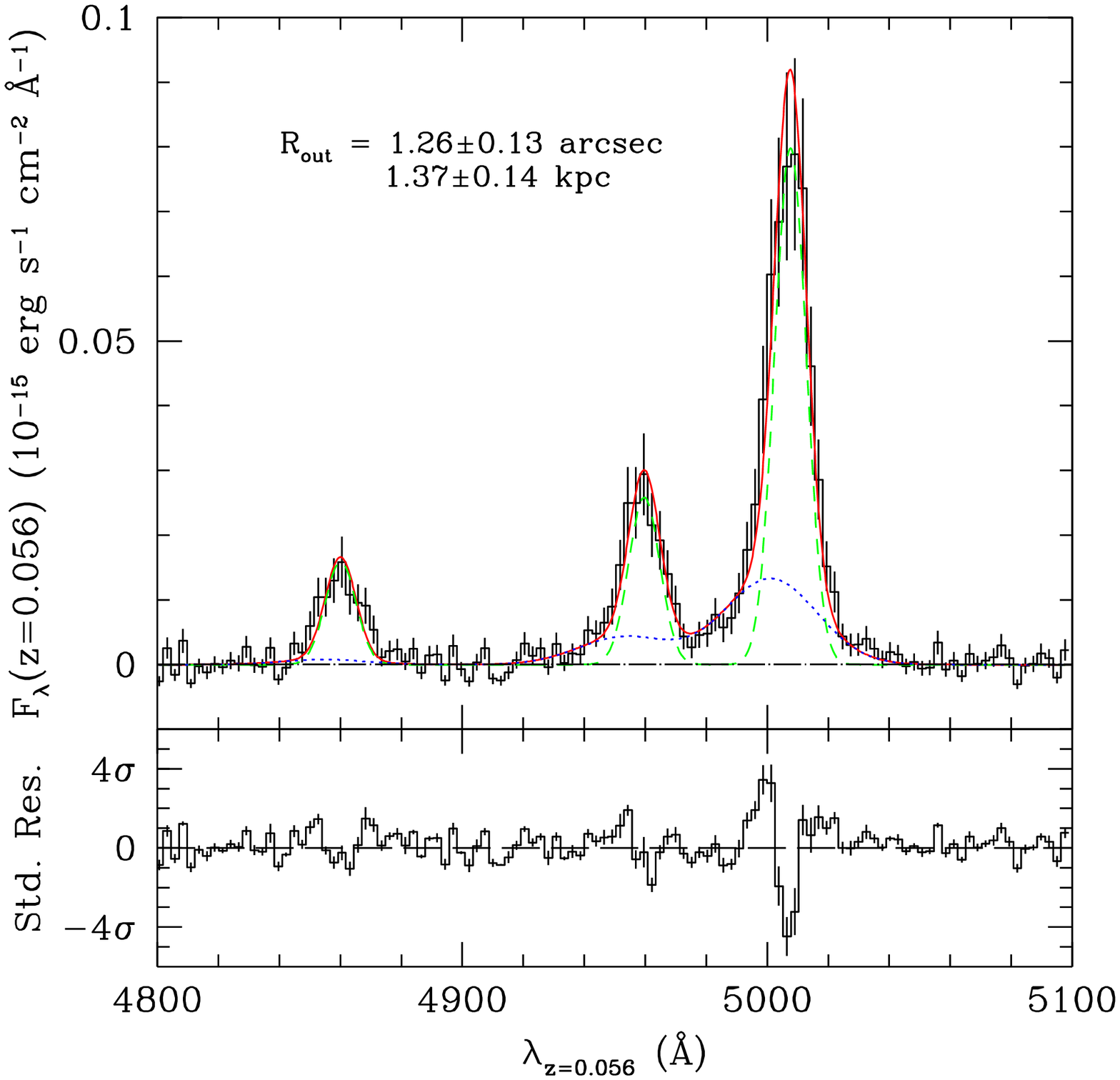}
\end{minipage}
\begin{minipage}{.49\textwidth}
\includegraphics[width=\textwidth]{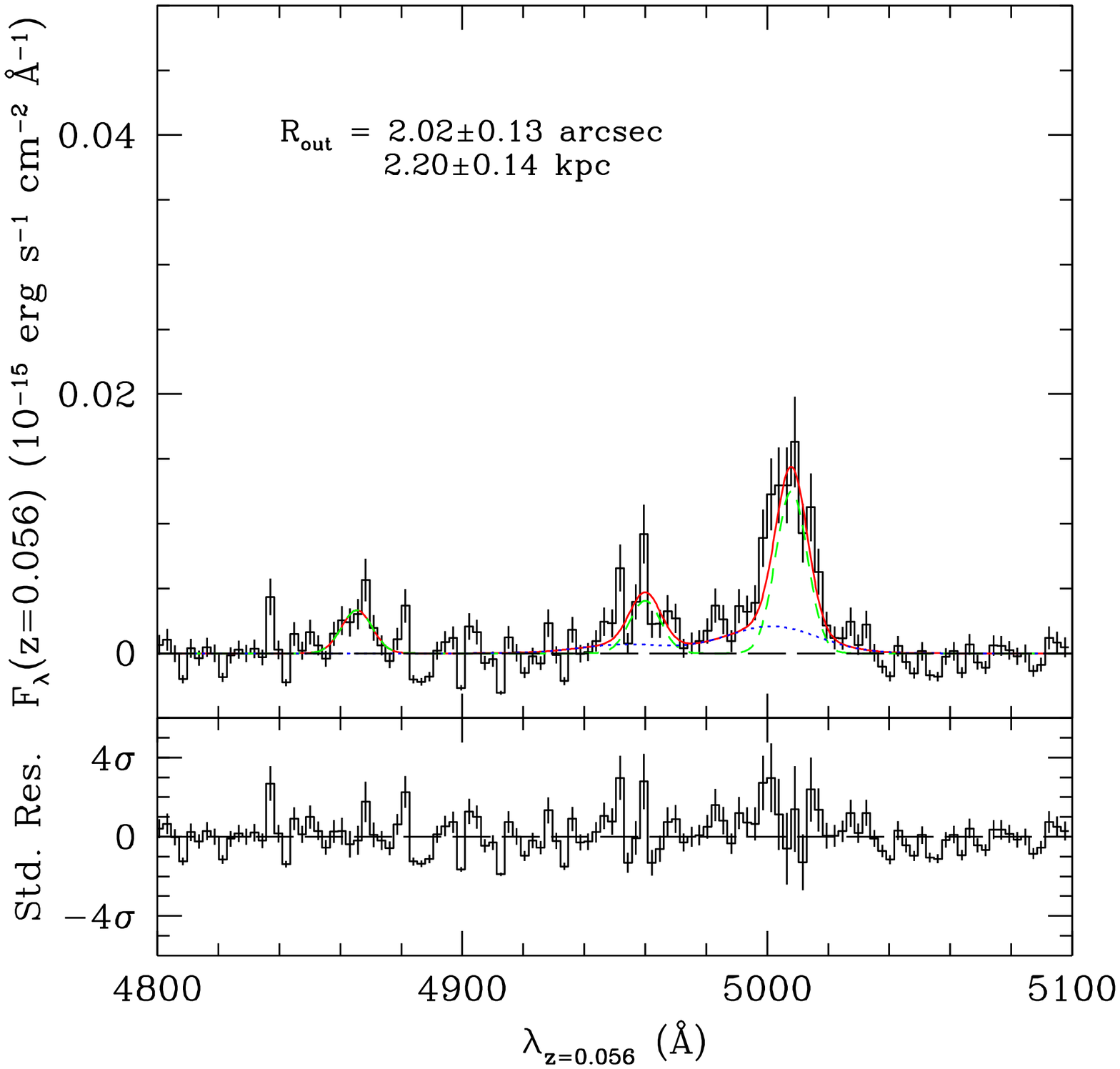}
\end{minipage}
\caption{Continuum-subtracted I20210S off-axis \hb+\oiii\ spectra extracted at a 5-px offset in the northern direction ({\itshape left panel}) and at an 8-px offset in the southern direction ({\itshape right panel}). In each panel, the best-fit model is shown (red solid line) along with the profiles of the narrow (green short-dashed line) and broad components (blue dotted line), and the zero-flux level is indicated (black long-dashed line). The standardized residuals after the best-fit subtraction are also shown in separate windows below each spectral region. The fit to the \hb\ emission is not accounted for the calculations of $\chi^2$ and $p_F$, which are performed on the \oiii\ doublet only (see text), and is shown here for visual purposes only.}
\label{fig:offaxis}
\end{figure*}

The electron density $\langle n_e \rangle$ is then related to the ratio $I_{6717}/I_{6731}$ between the components of the \s2\ doublet through:
\begin{equation}\label{eqn:dens}
\frac{I_{6717}}{I_{6731}} = 1.49 \cdot \frac{1 + 3.77 x}{1 + 12.8 x}
,\end{equation}
with $x = 10^{-2} \langle n_e \rangle T_e^{-1/2}$ \citep{Wee68,Ost06,San16}. However, we compute a ratio $I_{6717}/I_{6731} = 0.60 \pm 0.25$ for the ionized wind which is on the saturating side of Eq. \ref{eqn:dens}, and it only allows us to establish a lower limit at 95\% probability of $\langle n_e \rangle \gtrsim 4000$ cm$^{-3}$ to the outflow electron density. This might either be an indication of a high electron density or just a consequence of the severe blending that affects the \s2\ region at the low spectral resolution of DOLoRes, preventing us from deriving a reliable estimate of $\langle n_e \rangle$. The same issue holds for the \oii\ doublet, which could have been used in place of the \s2\ for such a measurement \citep{Ost06} but is even more blended because of its peak separation of $\sim$3 \AA\ only.
%This is likely due to the severe blending that affects the \s2\ region at the DOLoRes low spectral resolution, preventing us to derive a reliable estimate of $\langle n_e \rangle$.

As an alternative possibility for deriving solid estimates of $\langle n_e \rangle$ for the I20210S outflow, we also consider the application of the trans-auroral ratio (TR) method by \citet{Ros18}. This method, based on the evaluation of the line ratios -- \s2$_{4068,4076}$/\s2$_{6717,6731}$ and \oii$_{3726,3729}$/\oii$_{7319,7331}$ -- allows us to obtain at once both $\langle n_e \rangle$ and the intrinsic reddening $E(B-V)$ of the emitting gas. We thus fit the trans-auroral doublets \s2\ $\lambda\lambda$4068,4076 and \oii\ $\lambda\lambda$7319,7331 through the MC procedure with two narrow and two broad components each, fixing their widths to the corresponding values for the \oiii\ $\lambda\lambda$4959,5007 (see Table \ref{tab:pars}). This yields ${\rm TR}(\mbox{\s2})_{\rm out} = 0.192 \pm 0.051$ and ${\rm TR}(\mbox{\oii})_{\rm out} = 1.72 \pm 0.94$. Having derived for the outflow a ionization parameters $\log{U}_{\rm out} = -3.09 \pm 0.47$ from its relation to the \oiii/\hb\ and \n2/\ha\ ratios \citep[][BM19 hereafter]{Bar19}, we can finally compare its TRs to the simulations presented in \citet[][see their figure 7]{Dav20}, obtaining $\langle n_e \rangle_{\rm out} = 10,400^{+4000}_{-5200}$ cm$^{-3}$ and $E(B-V)_{\rm out} = 0.34^{+0.24}_{-0.15}$.

\begin{table}[htbp]
\centering
\resizebox{\columnwidth}{!}{
{\setstretch{1.25}
\begin{tabular}{lccr}
\hline
\hline
\multicolumn{4}{l}{ }\\
Quantity & \multicolumn{2}{c}{Value} & Units\\
\multicolumn{4}{l}{ }\\
\cline{2-3}
\multicolumn{4}{l}{ }\\
 & Outflow & South Nebula & \\
\multicolumn{4}{l}{ }\\
\hline
\multicolumn{4}{l}{ }\\
$T_e$ & $12,900 \pm 3600$ & $\sim$10,000 & K\\
$\langle n_e \rangle$ & $\gtrsim$5000 & $\sim$100 & cm$^{-3}$\\
$L_{\rm [OIII]}^{\rm out}$ & $(2.44 \pm 0.74) \times 10^{42}$ & $(9.05 \pm 0.55) \times 10^{39}$ & erg s$^{-1}$\\
$v_{\rm max}$ & $2160 \pm 380$ & $1100 \pm 430$ & km s$^{-1}$\\
$R_{\rm out}$ & $2.20 \pm 0.14$ & $6.52 \pm 0.43$ & kpc\\
$t_{\rm dyn}$ & $0.99 \pm 0.27$ & $5.8 \pm 2.6$ & Myr\\
\multicolumn{4}{l}{ }\\
\hline
\multicolumn{4}{l}{ }\\
$M_{\rm out}$ & $\left( 1.94^{+0.69}_{-0.51} \right) \times 10^5$ & $\sim 3 \times 10^{4}$ & M$_\odot$\\
$\dot{M}$ & $0.59^{+0.46}_{-0.26}$ & $\sim 6 \times 10^{-3}$ & M$_\odot$ yr$^{-1}$\\
$\dot{E}_{\rm kin}$ & $\left( 0.86^{+1.27}_{-0.54} \right)\times 10^{42}$ & $\sim 2 \times 10^{39}$ & erg s$^{-1}$\\
$\dot{P}_{\rm out}$ & $\left( 0.80^{+0.88}_{-0.44} \right) \times 10^{34}$ & $\sim 4 \times 10^{31}$ & erg cm$^{-1}$\\
\multicolumn{4}{l}{ }\\
\hline
\end{tabular}
}
}
\caption{Summary of the relevant physical properties of the ionized outflow discovered in the I20210S optical spectrum and of the South Nebula. {\itshape Upper section:} Quantities that are independent of the electron density. {\itshape Lower section:} Quantities dependent on the electron density, for which a value of $\langle n_e \rangle = 5000$ cm$^{-3}$ \citep{Ros18} is assumed in the case of the outflow. Note that the quoted errors of measurement are only indicative of the magnitude of the statistical uncertainties, not the systematics, affecting the computed values (see Sect. \ref{sec:disc}).}
\label{tab:outp}
\end{table}

As pointed out in the literature \citep{Ros18,Spe18,Dav20}, the TR method allows us to probe denser gas with respect to the use of the ``traditional'' \s2\ doublet, whose emission is likely produced at the ionization front where the electron density significantly decreases. This issue is probably at the base of the high values of ionized gas mass and mass outflow rate recently found in AGN winds \citep[e.g.,][]{Car15,Kak16,Bis17,Per17b}, for which values of $10^2$ cm$^{-3}$ $\lesssim \langle n_e \rangle \lesssim 10^3$ cm$^{-3}$ are usually assumed. Such an assumption is justified from measurements of the outflow electron density based on the \s2\ method: for example, \citet{Arr14} get $\langle n_e \rangle \sim 400$ cm$^{-3}$ for the outflowing emission in ULIRGs, whereas \citet{Per20} find $\sim$200 cm$^{-3}$ in the archetypal ULIRG Arp 220. For comparison, the values of $\langle n_e \rangle$ found by \citet{Ros18} for AGN-driven outflows in ULIRGs fall in the range $3000 \div 56,000$ cm$^{-3}$, with a median value of $\sim$5000 cm$^{-3}$. Also, \citet{Kak18} obtained spatially resolved values of $\langle n_e \rangle$ up to $\sim$2000 cm$^{-3}$ for ionized winds in nearby radio-selected Seyfert galaxies. Due to the limited DOLoRes spectral resolution and the severe blending that affects the I20210S trans-auroral emission lines with nearby features (e.g., the H$\delta$ close to the \s2\ $\lambda\lambda$4068,4076, the He I blueward and the [Ni {\scriptsize II}] redward of the \oii\ $\lambda\lambda$7319,7331), we cannot draw any firm conclusion on the reliability of the I20210S outflow electron density derived with the TR method. Therefore, in the following discussion of the physical properties of the I20210S ionized wind, we adopted $\langle n_e \rangle \sim 5000$ cm$^{-3}$ \citep{Ros18} as our main reference when computing all the related quantities.

With $L^{\rm out}_{\rm [OIII]} = (2.44 \pm 0.74) \times 10^{42}$ erg s$^{-1}$ obtained from the outflow \oiii\ flux reported in Table \ref{tab:pars}, and the further assumptions of $C \approx 1$ and ${\rm [O/H]} \sim {\rm [O/H]}_\odot$ (i.e., solar metallicity), Eq. \ref{eqn:mout} yields $M_{\rm out} = \left(1.94^{+0.69}_{-0.51}\right) \times 10^5$ M$_\odot$. Clearly, this value and those based on it are affected by the assumption on $\langle n_e \rangle$. We then derive the expression of the outflowing mass rate $\dot{M}$ from the fluid-field continuity equation as done in \citet{Bis17}, in order to provide a local estimate of this quantity at the outflow termination radius $R_{\rm out}$ \citep[e.g.,][]{Fer15}:
\begin{equation}\label{eqn:mdot}
\dot{M} = 3 \frac{M_{\rm out} v_{\rm max}}{R_{\rm out}}
.\end{equation}
In order to estimate the spatial extension of the outflow, we performed a series of adjacent, 1-px wide (i.e., $\sim$0.27 kpc, owing to the DOLoRes angular scale of $0.252$ arcsec px$^{-1}$ and the scale distance of 1.087 kpc arcsec$^{-1}$ at $z = 0.056$) extractions of the I20210S spectrum along its 2D trace in the high-S/N region of \hb+\oiii\ .

First, we fit a Gaussian function to the trace profile at the \oiii\ peak to get the trace width $\sigma = 1.67 \pm 0.01$ px (i.e., $\sim$0.46 kpc). Then, we extracted 1D off-axis spectra in both the northern and southern direction offset by 3 to 12 px ($\sim$0.8 to $\sim$3.3 kpc) from the aperture center \citep{Per15a,Bis17} in order to both exclude the signal enclosed in the instrumental PSF of $0''.85$ ($0''.43$ in each direction, i.e., 2.5 px) and avoid overlap with either I20210N or the South Nebula. Finally, we calibrated such spectra with the same wavelength dispersion and sensitivity function applied to the I20210S average spectrum, and applied the MC fitting procedure to the continuum-subtracted \oiii\ emission, only letting the line amplitudes free to vary -- narrow FWHM, broad FWHM and blueshifts are fixed to the values reported in Table \ref{tab:pars}. We produced the MC fit for both the two-component model and a comparison single-component model of the emission lines, in which the broad emission from the outflow is neglected. In this way, we are able to identify through a statistical $F$-test the transition region where the outflow signal becomes negligible with respect to the NLR emission: specifically, we define the significance threshold of the outflow by requesting an $F$-test probability $p_F > 0.90$.

The statistical analysis yields significant emission associated with the outflow up to 5 px ($\sim$1.3 kpc) in the northern direction, with $p_F \gtrsim 0.93$ ($\chi^2/\nu_{\rm d.o.f} = 76/73$); at larger distances along this direction, the signal emitted from the ionized wind quickly becomes indistinguishable from the noise, and thus has no impact on the best fit ($p_F = 0$). Instead, the outflow emission in the southern direction remains significant out to 8 px ($\sim$2 kpc, $p_F \gtrsim 0.93$, $\chi^2/\nu_{\rm d.o.f} = 78/73$). The ``terminal'' line spectra extracted at 5-px northward and 8-px southward offset are shown in Fig. \ref{fig:offaxis}. From this point on, we therefore adopt the distance $R_{\rm out} = 2.20 \pm 0.14$ kpc (i.e. $8.0 \pm 0.5$ px) as our fiducial value for the termination radius of the I20210S ionized wind within the applicability limits of the $F$-test statistics. This choice is motivated by the observation that an 8-px offset produces an extraction lying beyond 3$\sigma$ pixels from the trace center, and thus fiducially outside the 2.5-px PSF radius.

Then we calculate $v_{\rm max} = |\Delta v|_{\rm [OIII]}^{\rm out} + 2 \sigma_{\rm [OIII]}^{\rm out} = 2160 \pm 380$ km s$^{-1}$ \citep[see][and references therein]{Bis17} from the outflowing \oiii\ $\lambda$5007 parameters reported in Table \ref{tab:pars}. In this way, from Eq. \ref{eqn:mdot}, we obtain $\dot{M} = 0.59^{+0.46}_{-0.26}$ M$_\odot$ yr$^{-1}$. Finally, we derive the outflow kinetic power $\dot{E}_{\rm kin}$, the dynamical time scale $t_{\rm dyn}$ and the outflow momentum rate $\dot{P}_{\rm out}$ as:
\begin{eqnarray}\label{eqn:ekin}
\dot{E}_{\rm kin} = \frac{1}{2} \dot{M} v_{\rm max}^2,\\
t_{\rm dyn} \approx \frac{R_{\rm out}}{v_{\rm max}},\\
\dot{P}_{\rm out} = \dot{M} v_{\rm max}
,\end{eqnarray}
which yield $\dot{E}_{\rm kin} = \left( 0.86^{+1.27}_{-0.54} \right) \times 10^{42}$ erg s$^{-1}$, $t_{\rm dyn} = 0.99 \pm 0.27$ Myr and $\dot{P}_{\rm out} = \left( 0.80^{+0.88}_{-0.44} \right) \times 10^{34}$ erg cm$^{-1}$, respectively. We report all these quantities in Table \ref{tab:outp}, highlighting that, given the high reference value of $\sim$5000 cm$^{-3}$ adopted for the outflow $\langle n_e \rangle$, the electron-density dependent parameters might be underestimated by a factor of $\sim$10$\div$50.

It should also be noted that although the quantities presented in Table \ref{tab:outp} are given along with the errors, these should actually be interpreted as rough estimates, since the systematic uncertainties insisting on Eq. \ref{eqn:mout} exert a bias on them at the level of $1 \div 2$ orders of magnitude \citep{Bis17}. Given this caveat, the values of $M_{\rm out}$, $\dot{M}$ and $\dot{E}_{\rm kin}$ obtained for the outflow of I20210S are in line with those found by \citet{Rup13} for a sample of nearby galaxy mergers \citep[see also][]{Rod13,Spe18}. In order to definitively assess the AGN nature of the I20210S outflow, we compare its kinetic power to the expected ejected mass rate, $\dot{M}_{\rm SN}$, energy output, $\dot{E}_{\rm SN}$, and momentum injection, $\dot{P}_{\rm SN}$, of starbursts associated with supernova (SN) explosions \citep{Bru15}. According to \citet{Vei05}, such quantities are related to the host galaxy's star formation rate (SFR) by:
\begin{equation}\label{eqn:mdsn}
    \dot{M}_{\rm SN} \lesssim 0.26 \left(
    \frac{{\rm SFR}}{{\rm M}_\odot\mbox{ }{\rm yr}^{-1}}
    \right)
,\end{equation}
\begin{equation}\label{eqn:edsn}
    \dot{E}_{\rm SN} \lesssim 7 \times 10^{41} \left(
    \frac{{\rm SFR}}{{\rm M}_\odot\mbox{ }{\rm yr}^{-1}}
    \right)
,\end{equation}
\begin{equation}\label{eqn:pdsn}
    \dot{P}_{\rm SN} \lesssim 5 \times 10^{33} \left(
    \frac{{\rm SFR}}{{\rm M}_\odot\mbox{ }{\rm yr}^{-1}}
    \right)
,\end{equation}
whereas the SFR is linked to the host-galaxy IR ($8 \div 1000$ $\mu$m) luminosity $L^*_{\rm IR}$ by \citep{Ken98,Ken12,Ken21}:
\begin{equation}\label{eqn:sfr}
    \frac{{\rm SFR}}{{\rm M}_\odot\mbox{ }{\rm yr}^{-1}} = 3.9\times 10^{-44} \left(
    \frac{L^*_{\rm IR}}{{\rm erg}\mbox{ }{\rm s}^{-1}}
    \right)
.\end{equation}

We estimate $L^*_{\rm IR} \sim 3.4 \times 10^{44}$ erg s$^{-1}$ for I20210S from the values for the total IR luminosity of the I20210 system and the AGN IR luminosity for the single members presented in \citet{Ima14}, who estimated the contributions to the total galaxy IR emission coming from the active nucleus through photometric aperture size at high spatial resolution (see their Tables 1, 3, and 5). This in turn yields ${\rm SFR} \sim 13$ M$_\odot$ yr$^{-1}$, and hence $\dot{M}_{\rm SN} \lesssim 3.4$ M$_\odot$ yr$^{-1}$, $\dot{E}_{\rm SN} \lesssim 9 \times 10^{42}$ erg s$^{-1}$ and $\dot{P}_{\rm SN} \lesssim 6.5 \times 10^{34}$ erg cm$^{-1}$. Such values are $\sim$6 to $\sim$10 times higher than those listed in Table \ref{tab:outp}. Therefore, a starburst at work in I20210S is potentially able to produce the observed ionized outflow. However, \citet{Vei05} note that Eqs. \ref{eqn:mdsn} to \ref{eqn:pdsn} give the limit values for a thermalization efficiency of 100\% -- that is, when none of the starburst-injected energy is radiated away. Since typical starburst thermalization efficiencies are of the order of $\sim$10\% \citep[see][and references therein]{Vei05}, the actual energy output from SNe is expected to be (at most) in line with the values listed in Table \ref{tab:outp}. This fact, in combination with an IR emission powered by the AGN \citep{Ima14}, leads us to conclude that the wind in I20210S is likely AGN-driven, although a non-negligible contribution from star formation cannot be ruled out. Establishing the main driving mechanism of the I20210S outflow is even further challenged by the uncertainty in its electron density, which biases the derivation of the physical properties that can be directly compared with the expected SN energetics; future spatially resolved observations of I20210S will therefore also be of paramount importance in precisely assessing the nature of its ionized wind.

\begin{figure}[htbp]
\begin{center}
    \includegraphics*[width=.45\textwidth,angle=90]{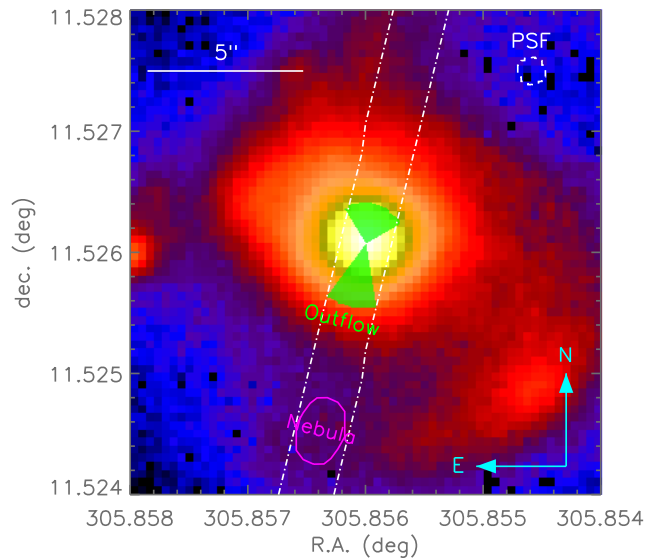}
\end{center}
\caption{Schematic representation of the emitting components observed in the spectrum of I20210S: the high-velocity outflow (green shaded area) -- assumed biconical for simplicity, since only one slit position is available and the PSF size prevents us from resolving the morphology of such a small structure -- and the South Nebula (magenta solid ellipse), plotted on top of the PS1 {\itshape grizy} image of the galaxy. The slit position and orientation (white dot-dashed lines) are reported; in the upper left corner, the diameter of $0''.85$ of the DOLoRes PSF (white dashed circle) is also indicated.}
\label{fig:cartoon}
\end{figure}

Broadened emission lines in ULIRGs hosting Seyfert nuclei are a common feature. \citet{Rod13} reported that up to 94\% of nearby objects of this kind ($z < 0.175$) show strongly blueshifted ($\Delta v > 150$ km s$^{-1}$) \oiii\ broad emission components (${\rm FWHM} > 500$ km s$^{-1}$) that are emitted by near-nuclear ($R_{\rm out} \lesssim 3.5$ kpc) warm ionized outflows. At the same time, while they are fully detectable in optical and UV spectra, such outflows are usually not capable of injecting enough power into the  surrounding environment of AGN to effectively affect the host-galaxy ISM and star formation. In the face of a required $\dot{E}_{\rm kin}$ of the order of 0.5\% to 5\% of the total AGN radiant energy \citep{DiM05,Hop10}, \citet{Fio17} showed, in fact, that the majority of near-nuclear warm outflows clusters around $\dot{E}_{\rm kin} \sim 0.001 L_{\rm bol}$ (see their Figure 1). The I20210S outflow appears to be consistent with this scenario, given its $\dot{E}_{\rm kin}/L_{\rm bol}$ ratio of $\sim$0.002. However, its power can still be sufficient to locally affect the star formation rate in some host-galaxy regions, as demonstrated by the anti-correlation found between the distribution of star-forming clouds and wind zones in AGN hosts over resolved spatial regions that are $\sim$3$\div$7-kpc wide \citep{Can12,Car15,Cre15}. Given its proximity, brightness, and spatial structure, I20210 therefore stands as an extremely peculiar laboratory in which the impact and interplay of ongoing galaxy merging on both star formation and AGN activity could be investigated in great detail.

\section{Physical properties of the South Nebula}\label{sec:detneb}

In this section, we briefly discuss the properties of the South Nebula. As described in Sect. \ref{sec:nebula}, this region exhibits interesting intermediate ionization properties between AGN-powered (FWHM $\sim 700$ km s$^{-1}$, velocity blueshift of $\sim$500 km s$^{-1}$) and star-forming gas clouds. We present the schematic structure of I20210S in Fig. \ref{fig:cartoon}, overlapping the position and extension -- within the spectrograph slit -- of both the outflow and the South Nebula to the PS1 image of the galaxy. From this picture, it is evident that the outflow propagating southwards extends outside enough of the innermost nuclear region to touch the inner regions of the disk structure, potentially interacting with the I20210S baryonic reservoir. Furthermore, the South Nebula is located on both the extensions of the outflow and the I20210S Western spiral arm, which makes it additionally interesting.

According to \citet{Hec00}, a galaxy with $L_{\rm IR} \sim 3 \times 10^{45}$ erg s$^{-1}$ as I20210S \citep{Pic10} has an average rotational velocity $\langle v_{\rm rot} \rangle$ of $\sim$250 km s$^{-1}$; at a distance of $\sim$2.2 kpc from the central engine, this translates to an escape velocity $v_{\rm esc} \sim 550$ km s$^{-1}$ when assuming a galactic radius of $\sim$6.5 kpc (i.e., the distance of the South Nebula). For comparison, the wind has a maximum velocity that is about four times higher (see Table \ref{tab:outp}), and the South Nebula itself exhibits $v_{\rm max} = 1100 \pm 430$ km s$^{-1}$ (about twice as high). Therefore, this implies that the nebula is being ejected outside the host galaxy by the ionized outflow, which also triggers possible star formation activity via quasar feedback as suggested by the placement of the South Nebula line ratios on the AGN-star formation boundary (see Sect. \ref{sec:classs}).

It is also interesting to consider its physical properties, as done in Sect. \ref{sec:disc} for the main outflow. To this end, we assume $T_e \sim 10^4$ K, which according to Eq. \ref{eqn:dens} translates to $\langle n_e \rangle \sim 100$ cm$^{-3}$ for the South Nebula \s2\ ratio $I_{6717}/I_{6731} = 1.36 \pm 0.10$ (see Tab \ref{tab:pars}). This, in turn, yields $M_{\rm out} \sim 3 \times 10^4$ M$_\odot$, given $L_{\rm [OIII]} = (9.05 \pm 0.55) \times 10^{39}$ erg s$^{-1}$ from the \oiii\ flux listed in Table \ref{tab:pars}, and finally $\dot{M} = M_{\rm out} v_{\rm max} / R_{\rm out} \sim 6 \times 10^{-3}$ M$_\odot$ yr$^{-1}$ \citep{Bis17} for $R_{\rm out} = 6.52 \pm 0.43$ kpc (see Fig. \ref{fig:blobs}). We report all of these quantities in Table \ref{tab:outp} along with the corresponding $t_{\rm dyn}$, $\dot{E}_{\rm kin}$ and $\dot{P}_{\rm out}$, to allow for a direct comparison with the values that hold for the outflow.

\section{I20210 SMBH mass estimates}\label{sec:mbhest}

In order to evaluate the SMBH mass $M_{\rm BH}$ in both objects, we used the approach detailed in BM19 for Type II AGN, in which obscuration prevents us from detecting the broad components of permitted emission lines. In this case, the following single-epoch relation linking $M_{\rm BH}$ to a BLR virial shape factor $\varepsilon$ that summarizes the uncertainities on the real BLR geometry, the monochromatic AGN luminosity $\lambda L_\lambda(5100\mbox{ \AA})$ at 5100 \AA\ and the broad \ha\ FWHM gives:
\begin{equation}\label{eqn:sembh}
\resizebox{0.9\hsize}{!}{$
\log{\left(
\frac{M_{\rm BH}}{{\rm M}_\odot}
\right)} = \log{\varepsilon} + 6.90 +0.54 \cdot \log{\left[
\frac{\lambda L_\lambda(5100\mbox{ }\AA)}{10^{44}\mbox{ }{\rm erg}\mbox{ }{\rm s}^{-1}}
\right]} + 2.06 \cdot \log{\left[
\frac{{\rm FWHM}^{\rm (BLR)}_{{\rm H}\alpha}}{10^3\mbox{ }{\rm km}\mbox{ }{\rm s}^{-1}}
\right]}
$}
\end{equation}
The validity of this relation holds as long FWHM$^{\rm (BLR)}_{{\rm H}\alpha}$ and the \oiii/\hb\ ratio are measured for AGN-dominated systems and are therefore related by the following logarithmic linear relation:
\begin{equation}\label{eqn:blrha}
\resizebox{0.9\hsize}{!}{$
\log{\left(
\frac{{\rm [O\mbox{ {\scriptsize III}}]}}{{\rm H}\beta}
\right)} = (0.58 \pm 0.07) \cdot \log{\left[
\frac{{\rm FWHM}^{\rm (BLR)}_{{\rm H}\alpha}}{{\rm km}\mbox{ }{\rm s}^{-1}}
\right]} - (1.38 \pm 0.38)
$}
\end{equation}
According to Figs. 3 and 4 of BM19, this happens for AGN-dominated systems with \oiii/\hb\ $\gtrsim 0.55,$ where the line intensities are not contaminated by star formation in the host galaxy. Since, based on Table \ref{tab:pars}, we have $\log{(\mbox{\oiii}/\mbox{\hb})} \sim 0.7$ for I20210N and $\sim$0.8 for I20210S, respectively, we can apply Eqs. \ref{eqn:sembh} and \ref{eqn:blrha} to both I20210 members. We do not quote any errors for the following estimations of physical quantities involved in the determination of $M_{\rm BH}$, since the measurement method is indirect and is therefore subject to uncertainties of at least $\sim$0.5 dex (see BM19 and references therein).

\begin{table}[htbp]
\centering
\resizebox{\columnwidth}{!}{
\begin{tabular}{lccr}
\hline
\hline
\multicolumn{4}{l}{ }\\
Quantity & I20210N & I20210S & Units\\
\multicolumn{4}{l}{ }\\
\hline
\multicolumn{4}{l}{ }\\
$L^{\rm (NLR)}_{{\rm H}\beta}$ & $(1.622 \pm 0.075) \times 10^{40}$ & $(1.040 \pm 0.072) \times 10^{42}$ & erg s$^{-1}$\\
$L_{\rm bol}$ & $5.2 \times 10^{43}$ & $3.6 \times 10^{45}$ & erg s$^{-1}$\\
$\lambda L_\lambda(5100\mbox{ }\AA)$ & $7.0 \times 10^{42}$ & $2.7 \times 10^{44}$ & erg s$^{-1}$\\
${\rm FWHM}^{\rm (BLR)}_{{\rm H}\alpha}$ & 3650 & 5700 & km s$^{-1}$\\
$M_{\rm BH}$ & $2.9 \times 10^7$ & $5.2 \times 10^8$ & M$_\odot$\\
$\lambda_{\rm Edd}$ & 0.01 & 0.05 & ---\\
$\sigma_v^*$ & $390 \pm 50$ & --- & km s$^{-1}$\\
$M_*$ & $\lesssim$1.5$\times 10^{12}$ & --- & M$_\odot$\\
\multicolumn{4}{l}{ }\\
\hline
\end{tabular}
}
\caption{Physical parameters of the AGN hosted in the I20210 members. Quantities derived from the application of proportionality relations (Eq. \ref{eqn:sembh} to Eq. \ref{eqn:lmon}) are reported without errors due to the uncertainties of at least $\sim$0.5 dex affecting them (see BM19 and references therein).}
\label{tab:mbhns}
\end{table}

For I20210N, we derive $\lambda L_\lambda(5100\mbox{ \AA})$ using its absorption-corrected hard X-ray luminosity $L_{2-10\mbox{ }{\rm keV}} = 4.7 \times 10^{42}$ erg s$^{-1}$ \citep{Pic10} via Eq. 5 from \citet{Mai07}:
\begin{equation}\label{eqn:x2opt}
\log{L_{2-10\mbox{ }{\rm keV}}} = 0.721 \cdot \log{\left[
\lambda L_\lambda(5100\mbox{ }\AA)
\right]} + 11.78.
\end{equation}
This yields $\lambda L_\lambda(5100\mbox{ }\AA) \simeq 7.0 \times 10^{42}$ erg s$^{-1}$. The FWHM of the invisible \ha\ broad component is estimated from Eq. \ref{eqn:blrha}; in this way, we find ${\rm FWHM}^{\rm (BLR)}_{{\rm H}\alpha} \simeq 3650$ km s$^{-1}$, corresponding to $M_{\rm BH} \simeq 2.9 \times 10^7$ M$_\odot$ if the value of $\varepsilon = 1.075$ by \citet{Rei15}, which is representative of assuming the mean virial factor $\langle f \rangle = 4\varepsilon = 4.3,$ derived by \citet{Gri13} by measuring the stellar velocity dispersion in the host galaxies of powerful nearby quasars.

For I20210S, we decided to avoid directly obtaining $\lambda L_\lambda(5100\mbox{ \AA})$ from the observed spectrum due to the uncertainty on its intrinsic reddening; similarly, we did not apply Eq. \ref{eqn:x2opt} to indirectly compute it, since only a lower limit to $L_{2-10\mbox{ }{\rm keV}} \gtrsim 5 \times 10^{43}$ erg s$^{-1}$ is reported in \citet{Pic10}. Instead, we relied on Eq. 6 from BM19:
\begin{equation}\label{eqn:lbol}
\resizebox{0.9\hsize}{!}{$
\log{L_{\rm bol}} = \log{L^{\rm (NLR)}_{{\rm H}\beta}} + 3.48 + \max{\left\{
0,\mbox{ }0.31 \cdot \left[
\log{\left(
\frac{{\rm [O\mbox{ {\scriptsize III}}]}}{{\rm H}\beta}
\right)} - 0.6
\right]
\right\}}
$}
\end{equation}
and Eq. 6 by \citet{Net09}:
\begin{equation}\label{eqn:lmon}
\log{\left[
\lambda L_\lambda(5100\mbox{ }\AA)
\right]} = 1.09 \cdot \log{L_{\rm bol}} - 5.23
.\end{equation}
From the value for I20210S of $L^{\rm (NLR)}_{{\rm H}\beta} = (1.040 \pm 0.072) \times 10^{42}$ erg s$^{-1}$ computed from the \hb\ flux listed in Table \ref{tab:pars}, we were thus able to derive $L_{\rm bol} \simeq 3.6 \times 10^{45}$ erg s$^{-1}$ and $\lambda L_\lambda(5100\mbox{ }\AA) \simeq 2.7 \times 10^{44}$ erg s$^{-1}$. The value for $L_{\rm bol}$ estimated in this way is fully compatible with the value of $\sim$3$\times 10^{45}$ erg s$^{-1}$ assumed by \citet{Pic10} on the basis of the I20210S infrared luminosity in the range $10 \div 100$ $\mu$m \citep{Sar11}. Finally, Eq. \ref{eqn:blrha} yields ${\rm FWHM}^{\rm (BLR)}_{{\rm H}\alpha} \simeq 5700$ km s$^{-1}$, which translates into $M_{\rm BH} \simeq 5.2 \times 10^8$ M$_\odot$.

These SMBH masses imply an Eddington luminosity $L_{\rm Edd} \sim 3.7 \times 10^{45}$ erg s$^{-1}$ for I20210N and $\sim$6.6$\times 10^{46}$ erg s$^{-1}$ for I20210S, respectively. Combining them with the AGN bolometric luminosities of $\sim$5.2$\times 10^{43}$ erg s$^{-1}$ for I20210N and $\sim$3.6$\times 10^{45}$ for I20210S, we obtain Eddington ratios $\lambda_{\rm Edd} = L_{\rm bol}/L_{\rm Edd} \sim 0.01$ for I20210N and $\sim$0.05 for I20210S. Although these values are affected by large uncertainties, such ratios are consistent with the galaxy classification from the BPT diagrams, with I20210S clearly falling in the Seyfert region and with I20210N exhibiting more mixed properties. In addition, we combine the system of equations in Eq. 9 from BM19 to infer the stellar mass $M_*$ of the I20210N host galaxy from its $\sigma_v^*$ of 390 km s$^{-1}$ (derived in Sect. \ref{sec:fitprocn}), finding $M_* \sim 1.5 \times 10^{12}$ M$_\odot$. We summarize all these parameters in Table \ref{tab:mbhns}.

The value of $M_*$ derived for I20210N is $\sim$50 times larger than that typically expected from the $M_{\rm bulge}$-to-$M_{\rm BH}$ relation \citep[$M_{\rm bulge} \simeq 10^3 M_{\rm BH}$; e.g.,][]{Mag98,Har04,Gul09}; however, we highlight that the I20210N kinematics is likely altered by its gravitational interaction with I20210S, thus making its measured $\sigma_v^*$ unreliable for the purposes of estimating the stellar mass. Therefore, the value of $M_*$ derived in this way should only be treated as an (overestimated) upper limit to the I20210N baryonic content.

\section{Summary and conclusions}\label{sec:conc}

In this article, we present an optical spectroscopic analysis of the AGN pair hosted in the interacting system IRAS 20210+1121 \citep[I20210; P90;][]{Hei95,Bur01,Dav02,Arr04,Pic10} at $z = 0.056$. This study is based on spectroscopy taken through a slit aligned along the nuclei of the two interacting galaxies. The high-quality data taken at the {\itshape Telescopio Nazionale Galileo} allowed us to perform a detailed study of the light emission from both components and from their surrounding environment in the rest-frame wavelength range of 3500 -- 7300 \AA, with the possibility of a comprehensive characterization of this interacting galaxy pair. Here we summarize our main findings:
\begin{itemize}
\item I20210N, the northern member of the I20210 system, can be definitively classified as a Seyfert 2 galaxy with an exceptional stellar velocity dispersion of $\sigma_v^* \sim 400$ km s$^{-1}$, hosting an AGN powered by a black hole with $M_{\rm BH} \sim 3 \times 10^7$ M$_\odot$ that radiates at 1\% of its Eddington limit.
\item The environment around I20210S, the southern component is a powerful Type II quasar with $M_{\rm BH} \sim 5 \times 10^8$ M$_\odot$ radiating at 5\% of its Eddington limit and revealed to be highly structured, with an ionized outflow and a detached gaseous nebula (the South Nebula) alongside the nuclear emission.
\item The physical properties of the ionized outflow derived from the analysis of the broad emission-line components ($T_e \sim 10^4$ K, $\langle n_e \rangle \gtrsim 5000$ cm$^{-3}$, $v_{\rm max} \sim 2000$ km s$^{-1}$, $R_{\rm out} \sim 2$ kpc, $M_{\rm out} \sim 2 \times 10^5$ M$_\odot$, $\dot{M} \sim 0.6$ M$_\odot$ yr$^{-1}$) are in line with those found in other powerful AGN hosted in ULIRGs \citep{Rod13,Rup13,Kak18,Spe18}. This suggests that the I20210S AGN activity has potentially a direct impact on the host-galaxy environment through quasar feedback; however, these results need to be further investigated with higher resolution spectral observations in order to constrain the value of the wind electron density and thus allow for a better characterization of the feedback mechanism to the  star formation activity in I20210S \citep[e.g.,][]{Car15,Fio17};
\item The South Nebula exhibits dynamical properties consistent with those of highly disrupted gas stripped out of the I20210S nucleus (velocity blueshift of $\sim$500 km s$^{-1}$, FWHM of $\sim$700 km s$^{-1}$ that is similar to the case of the Teacup Galaxy \citep{Ram17} and coupled to intermediate ionization properties between AGN-powered and star-forming gas. Such features qualify this region as a very interesting target for a deeper investigation of the potential feedback processes -- either triggered by AGN activity or by the galaxy merger -- at work in I20210S.
\end{itemize}

Thanks to the above properties, the I20210 system can be characterized as a very interesting target in the local Universe that ought to be investigated with dedicated multi-wavelength follow-ups aimed at a detailed study of the effects of AGN feedback coupled to host-galaxy interaction on the AGN surrounding environment. In particular, obtaining higher resolution spectra ($\lambda/\Delta\lambda \gtrsim 1500$) is crucial to improving the emission-line diagnostics of the I20210S components (nucleus, outflow, South Nebula) and to allow for a precise evaluation of the I20210S outflow physical conditions. Furthermore, integral-field spectroscopic observations are required to accurately constrain both the morphology and interplay of outflows and any off-nuclear emitting region in I20210S.

\begin{acknowledgements}
We thank our anonymous referee for their helpful comments. GV, EP, CV, MB, CF and FF acknowledge support from PRIN MIUR project ``Black Hole winds and the Baryon Life Cycle of Galaxies: the stone-guest at the galaxy evolution supper'', contract \#2017PH3WAT. GV also acknowledges financial support from Premiale 2015 MITic (PI: B. Garilli). CRA acknowledges financial support from the Spanish Ministry of Science, Innovation and Universities (MCIU) under grant with reference RYC-2014-15779, from the European Union's Horizon 2020 research and innovation programme under Marie Sk\l odowska-Curie grant agreement No 860744 (BiD4BESt), from the State Research Agency (AEI-MCINN) of the Spanish MCIU under grants "Feeding and feedback in active galaxies" with reference PID2019-106027GB-C42, "Feeding, feedback and obscuration in active galaxies" with reference AYA2016-76682-C3-2-P, and "Quantifying the impact of quasar feedback on galaxy evolution (QSOFEED)" with reference EUR2020-112266. CRA also acknowledges support from the  Consejería de Econom{\'i}a, Conocimiento y Empleo del Gobierno de Canarias and the European Regional Development Fund (ERDF) under grant with reference ProID2020010105 and from IAC project P/301404, financed by the Ministry of Science and Innovation, through the State Budget and by the Canary Islands Department of Economy, Knowledge and Employment, through the Regional Budget of the Autonomous Community. Based on observations made with the Italian {\itshape Telescopio Nazionale Galileo} ({\itshape TNG}) operated on the island of La Palma by the Fundaci{\'o}n Galileo Galilei of the INAF (Istituto Nazionale di Astrofisica) at the Spanish Observatorio del Roque de los Muchachos of the Instituto de Astrof{\'i}sica de Canarias. {\scriptsize IRAF} is distributed by the National Optical Astronomy Observatories, which is operated by the Association of Universities for Research in Astronomy, Inc. (AURA) under cooperative agreement with the National Science Foundation. Reproduced with permission from Astronomy \& Astrophysics, \textcopyright\ ESO.
\end{acknowledgements}

\bibliographystyle{aa}
\bibliography{biblio}

\end{document}